\documentclass[conference]{IEEEtran}
\IEEEoverridecommandlockouts

\usepackage{cite}
\usepackage{amsmath,amssymb,amsfonts}
\usepackage{amsthm}
\usepackage{algorithmic}
\usepackage{graphicx}
\usepackage{textcomp}
\usepackage{xcolor}
\def\BibTeX{{\rm B\kern-.05em{\sc i\kern-.025em b}\kern-.08em
    T\kern-.1667em\lower.7ex\hbox{E}\kern-.125emX}}
    
\usepackage{threeparttable}

\usepackage{tikz}
\usetikzlibrary{arrows.meta}

\usepackage{lscape}
\usepackage{hyperref}

\usepackage{mathtools}

\DeclarePairedDelimiter\floor{\lfloor}{\rfloor}

\usepackage{listings}
\usepackage{blindtext}
\usepackage[ruled,linesnumbered]{algorithm2e}
\usepackage{etoolbox,xstring,mfirstuc,textcase}
\usepackage{booktabs} 
\usepackage{listings}

\usepackage{comment}
\usepackage{subfigure} 

\usepackage{xcolor}
\newtheorem{clm}{Claim} 
\newtheorem{thm}{Theorem}    
\newtheorem{lem}{Lemma}  
\newtheorem{defi}{Definition}

\usepackage{fancyhdr} 
\newenvironment{prf}{{\noindent\it Proof\,\,}}{\hfill $\square$\par}
    
\usepackage[utf8x]{inputenc}

\begin{document}

\title{Formal Security Analysis on dBFT Protocol of NEO \\ \LARGE (Extended Version)$^\flat$\thanks{$^\flat$This work covers our previous observation in \cite{fc20neo}. In this paper, we further provide a formal analysis on the insecurity of dBFT (based on SMR model). We generic the problems in dBFT to avoid similar pitfalls. We also provide comprehensive discussions that relates to our identified issues. The work shows an exploration to a real-world open-sourced blockchain project. }}

\author{\IEEEauthorblockN{
Qin Wang \IEEEauthorrefmark{1}\IEEEauthorrefmark{4},
Rujia Li \IEEEauthorrefmark{2}\IEEEauthorrefmark{3},
Shiping Chen \IEEEauthorrefmark{4},
Yang Xiang \IEEEauthorrefmark{1}}
\IEEEauthorblockA{\IEEEauthorrefmark{1} \textit{Swinburne University of Technology}, Melbourne, Australia.\\
\IEEEauthorrefmark{2} \textit{Southern University of Science and Technology}, Shenzhen, China.\\
\IEEEauthorrefmark{3} \textit{University of Birmingham}, Birmingham, UK.\\
\IEEEauthorrefmark{4} \textit{CSIRO Data61}, Sydney, Australia. \\
}
}


\maketitle

\begin{abstract}
NEO is one of the top public chains worldwide. We focus on its backbone consensus protocol, called delegated Byzantine Fault Tolerance (dBFT). The dBFT protocol has been adopted by a variety of blockchain systems such as ONT. dBFT claims to guarantee the security when no more than $f =  \lfloor \frac{n}{3} \rfloor$ nodes are Byzantine, where $n$ is the total number of consensus participants. However, we identify attacks to break the claimed security. In this paper, we show our results by providing a security analysis on its dBFT protocol. First, we evaluate NEO’s source code and formally present the procedures of dBFT via the state machine replication (SMR) model.  Next, we provide a theoretical analysis with two example attacks. These attacks break the security of dBFT with no more than $f$ nodes. Then, we provide recommendations on how to fix the system against the identified attacks. The suggested fixes have been accepted by the NEO official team. Finally, we further discuss the reasons causing such issues, the relationship with current permissioned blockchain systems, and the scope of potential influence. 
\end{abstract}

\section{Introduction}
NEO \cite{NeoWP}\cite{Neo18} has become one of the famous blockchain platforms  \cite{ferrag2018blockchain}\cite{dai2019blockchain} prevailing in the world since 2017. At the same time, NEO was the earliest and the longest everlasting public chain platform in China\footnote{https: //coinmarketcap.com/zh/currencies/neo/historical-data/}. The market capitalization of NEO, at the time of witting, reached around 0.78 billion USD. As an open-source, community-driven platform, NEO provides a developer-friendly environment to realize digital projects. A matured ecology with varieties of decentralized applications (DApps) has been established, covering Games, Lotteries, Wallets, and Exchanges.  Developers across the world are active, forming one of the largest communities following Ethereum \cite{Eth14}. Furthermore, aiming at providing a smart economy system on the top of distributed networks, NEO has developed a complete architecture from basic components including its consensus mechanism and P2P networks, to upper-layer components such as NeoX, NeoFS, NeoQS  \cite{NeoWP}\cite{Neo18}. In this paper, we focus on its underlying consensus protocol, the delegated Byzantine Failure Tolerance (dBFT).

Consensus protocols allow distributed participants to collectively reach an agreement, in a way that all honest participants agree on the same decision. This enables the immutability of blockchains and prevents double-spending attacks \cite{nakamoto2008bitcoin}. Traditional consensus mechanisms, such as Byzantine fault tolerance (BFT), can tolerate a number of Byzantine nodes who perform arbitrary actions. BFT systems are permissioned in which the consensus committee consists of a small fixed number of nodes. The participated nodes provide a deterministic consensus guarantee \cite{vukolic2015quest} \cite{vukolic2017rethinking} for the agreement. BFT protocols have been adopted as the foundations of many variants and applied to various projects for special needs. For instance, a variant of Practical Byzantine Fault Tolerance (PBFT) \cite{CastroL99} has been implemented for Hyperledger Fabric v0.5/v0.6 \cite{hyperledgerfabric} and Hyperledger Sawtooth v1.0 \cite{hyperledgersawthooth}\cite{androulaki2018hyperledger}. The protocol enables systems to run efficiently (with polynomial complexity) and safely (tolerate participants with arbitrary faults). 

Similarly, dBFT is also a variant of PBFT. Three types of modifications in dBFT are identified compared to PBFT, including the architect model (from Client/Server to P2P), permission rules (from fixed to dynamic), and consensus procedures (from three-phase to two-phase). Based on such a design, dBFT provides a balance between performance and scalability. However, security has not been seriously evaluated. Performing a comprehensive analysis on its security is absent from official documents such as the whitepaper \cite{NeoWP}, website \cite{Neo18}, and GitHub \cite{NeoGit}. In fact, after digging into the source code, we observed that the implemented dBFT is different from the descriptions presented in its whitepaper. For instance, only a part of messages should be signed in the document \cite{NeoWP}, however, all transferred messages are signed in the actual implementation. For deep and thoughtful analyses concerning its security, in this paper, we provide the following \textit{contributions}.

\textit{\textbf{Review of dBFT.} }
We have analyzed the source code to review the entire dBFT protocol, which is based on the version of $\mathsf{git commit}$ $\mathsf{5df6c2f05220e57f4e3180dd23e58bb2f675457d}$. Generally speaking, three types of nodes are involved in dBFT, namely, \textit{speaker} (leader), \textit{delegate} (backups) and \textit{common} node (client). The speaker plays the role of a leader and makes up a committee together with delegates to run the consensus algorithm. Common nodes do not participate in consensus processes but only synchronize messages. dBFT is a two-phase (2PC) protocol that contains the phases of PREPARE and RESPONSE. A speaker among delegates is selected as the leader, and then the delegate replies to all participated members with his response on \textit{success} or \textit{failure}. After receiving more than threshold responses on proposals, the member decides the next step: move on, rollback, or change view. Additionally, we also examine other modifications, such as the network model, rule of permission, \textit{etc.} We omit them to narrow the paper scope.

\textit{\textbf{Establish Formal Model.} }
We provide a formal treatment of dBFT with the security goals in a clear and accurate way. The formal presentation of dBFT is based on State Machine Replication (SMR) model \cite{Schneider90}\cite{lynch1996}\cite{castro2001full}. A \textit{state} of dBFT is transited into a new state under the triggering of \textit{action}s. We provide the formalized network model, involved entities (client, speaker, backups), and major components (consensus, view change). This model assists in presenting the full view of the protocol from its source code. Based on that, we define the security model of dBFT adopted in BFT-style protocols, containing both \textit{safety} and \textit{liveness}. Typically, the safety is used to guarantee that the consensus cannot decide on two different values. The liveness ensures a decision would be eventually made. Only when these two properties are satisfied at the same time, we say that the dBFT protocol is secure. The formalization procedure provides a clear and convincible presentation for finding vulnerabilities, especially when the protocol specifications have not been fully released. 

\begin{figure}[!htbp]
\centering
\includegraphics[width=0.4\textwidth]{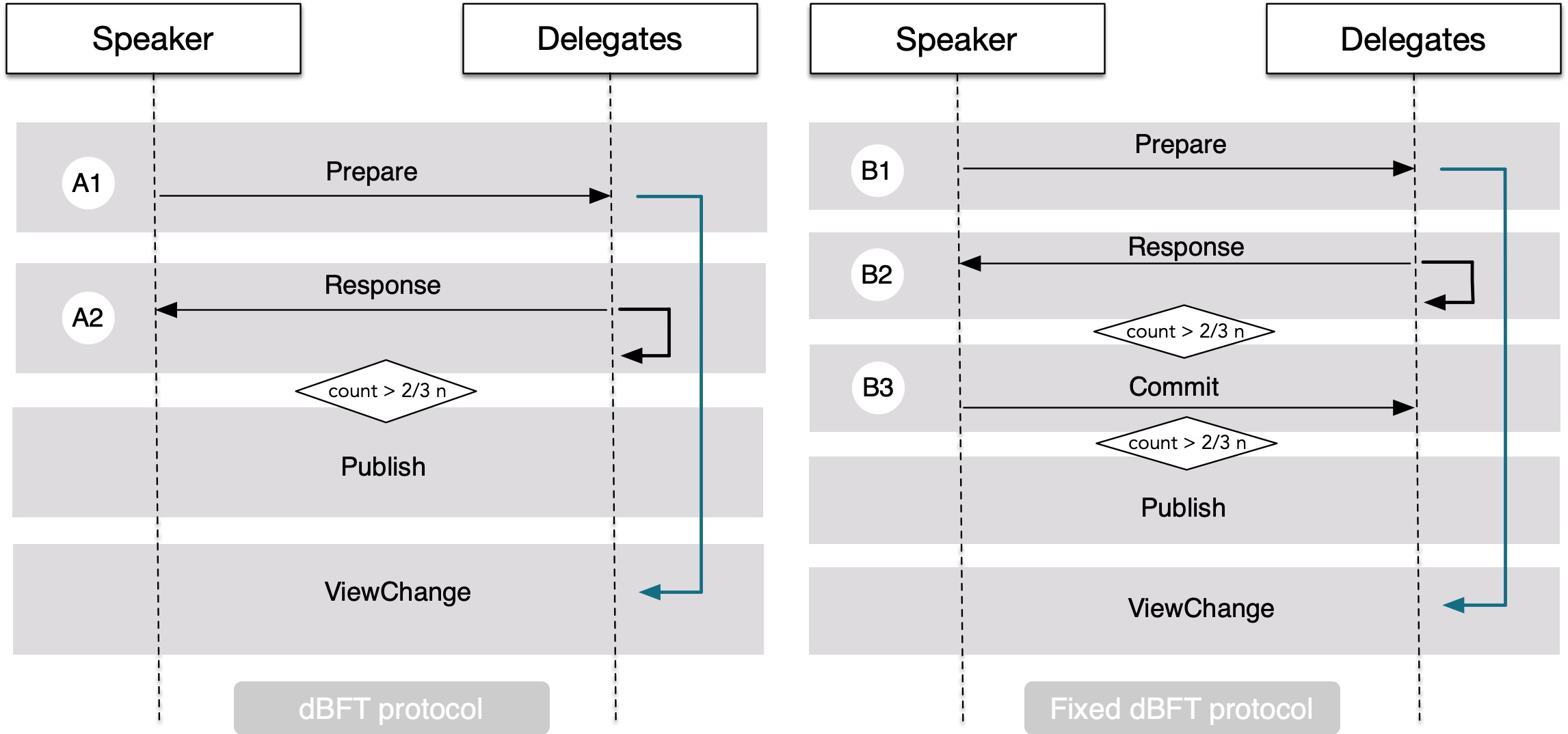}
\caption{Comparison between dBFT and Fixed dBFT
}
\label{3PC}
\end{figure}

\textit{\textbf{Security Analysis of dBFT.}}
We provide a strict security analysis under our formal model of dBFT. Based on our formal analysis, the dBFT protocol confronts the safety issue. To briefly show the issue, we present the current dBFT protocol in the left column in Fig.\ref{3PC} and denote each phase with the symbol $(A1, A2)$. Suppose a portion of nodes failed in phase $A2$, their states may not synchronize to peers due to certain reasons. In this situation, although each node knows his own decision, he still has no knowledge of others' decisions. When more than one conflicting decision is spread in the network, the consensus fails if no restart mechanism exists. dBFT employs the \textit{Viewchange} phase to reset the system. However, \textit{Viewchange} cannot guarantee the consistency caused by malicious attacks. For instance, compromised nodes in phase $A2$ may delay the message to split the network, making half of the consensus nodes accept decisions and ready to move on, while others do not. This breaks both the safety and liveness of the system. The former is due to the reason where inconsistent agreements make a coin spent multiple times. The latter is a side effect of the former --- if the situation of inconsistent states is not resolved, the system cannot proceed. A failed consensus will cause irrevocable financial loss \cite{Sam2021sok}.


\textit{\textbf{Attacks in dBFT.} } 
We have identified two types of attacks on dBFT. Both attacks are feasible with no more than  $\lfloor \frac{n}{3} \rfloor$ nodes. The first attack requires the speaker to be Byzantine, and the second attack requires the package delay in the network. We assume that there are a total of seven delegates in the system performing the consensus. Although the attacks can be launched with any number of delegates, we set seven committee nodes to stay aligned with the initial configurations of the NEO official team. According to the quorum theory \cite{vukolic2012quorum}, the system should be able to tolerate at most $f=\floor{\frac{n-1}{3}}$ Byzantine nodes under $n=7$. However, we provide the attacks in the case of $f=2$ (malicious speaker) and $f=1$ (package delay), respectively. We also provide concrete analyses based on these instantiations to show why dBFT is vulnerable to two or fewer malicious nodes. 

\textit{\textbf{Apply Fixes.} }
We further provide our recommended fixes for the identified problems and project has been practically fixed according to our support \cite{neo320}\cite{neo547}. The fix of the issue is straightforward. Adding the \textit{Commit} phase to the protocol, aiming to check whether enough (at least $2f+1$) delegates have responded to the request. If a node receives $2f+1$ signed responses from the delegates in the \textit{Response} phase, it commits the block by signing the response message together with state information. If receiving $2f+1$ commits messages, the node updates the local state of the blockchain by packing the block and broadcast the committed blocks to the network. As shown in Fig.\ref{3PC}, the fixed three-phase protocol splits the \textit{Response} phase $A2$ into two independent phases $B2$ and $B3$. The added phase $B3$, called \textit{Commit}, is used to decouple the task of $A2$. Specifically, in phase $B2$, the speaker and delegates can obtain the decisions from other consensus nodes with no need to publish the block immediately. After receiving enough signatures, they enter the \textit{Commit} phase, $B3$. In this phase, consensus nodes are aware of whether other committee nodes are ready to execute the same decisions (synchronize the same transactions/blocks in the next step) or not. If enough signatures (over threshold) get confirmed, the nodes can execute the PUBLISH process. In fact, this is a well-studied problem. It is possible to have a secure two-phase protocol for crash fault tolerance protocols (CFT), where the nodes can only be crashed. However, if in a system contains a group of Byzantine nodes who can behave arbitrarily, the additional \textit{Commit} phase becomes necessary  \cite{CastroL99}\cite{FischerLP85}. As a standard construction derived from classic BFT protocols, a three-phase solution is proved to be secure in \cite{RahliVVV18}.


\textit{\textbf{Further Discussion.} }
We present several related questions to extend the scope of the reported issue of this work. Correspondingly, we provide comprehensive answers based on sufficient existing studies. In the above context, we report the issue and perform a solution to repair it. However, it is not enough. The myths surrounding this reported issue needs to be theoretically answered, and further discussions on the similar issue are crucial and urgent, because dBFT has been widely adopted by several high-valued projects such as the Ontology \cite{Ont18}. We list several related questions from a general view, to explore why the same issue happens in many industrial instances. The topics include: (1) How does the issue come from; (2) Will the issue be thoroughly solved; (3) How to mitigate the issue;  (4) Will the issue impact other BFT-style permissioned blockchains; and (5) What is the lesson learned from this work. The corresponding answers are based on our formalized analysis and existing theories. We believe that this work is educational to the design and development of consensus protocols, and can be used as a reminder to avoid the same type of issues. In a nutshell, based on previous explanations, a short version of \textbf{\textit{contributions}} are:

%


\begin{itemize}
\item[-] We provide the first clear and accurate presentation of the dBFT protocol (in particular, for the NEO project \cite{Neo18}) based on its source code \cite{neomaster} $\mathsf{git\,commit}$  \\ $\mathsf{5df6c2f05220e57f4e3180dd23e58bb2f675457d}$.

\item[-]  We provide a formal treatment of the dBFT protocol by employing the State Machine Replication model (SMR) \cite{castro2001full}. The investigated source code version keeps pace with the version in \cite{fc20neo}.

\item[-] We conduct a rigorous analysis of the security of dBFT, and identify two attacks against the claimed security promise. We provide two concrete example attacks with visualized graphs derived from our formal analysis, to illustrate how they may work in practice.

\item[-] We have fully disclosed our results, covering both identified vulnerabilities and recommended fixes, to the NEO team. They acknowledge that the attacks are effective in their system, and the reported problems have been properly fixed with the help of our suggestions \cite{neo320}\cite{neo547}.

\item[-]  We discuss the origination of the issue and the relationship with current permissioned blockchain systems. We also discuss the potential impacts of such security issues.

\end{itemize}

The rest of our paper is structured as follows: A formal dBFT protocol based on the SMR model is presented in Section~\ref{sec-dbftview}. Formal security analysis is provided in Section \ref{sec-analysis}. Two example attacks with visualized workflow are shown in Section \ref{sec-example}. The discussions on identified security issues are listed in Section \ref{sec-discussion}. Related studies surrounding this paper are presented in \ref{sec-relatedwk}. Finally, the summary and future work are provided in Section \ref{sec-conclusion}.

\section{dBFT Formal Presentation}
\label{sec-dbftview}

\begin{figure}[!htbp]
\centering
\includegraphics[width=0.43\textwidth]{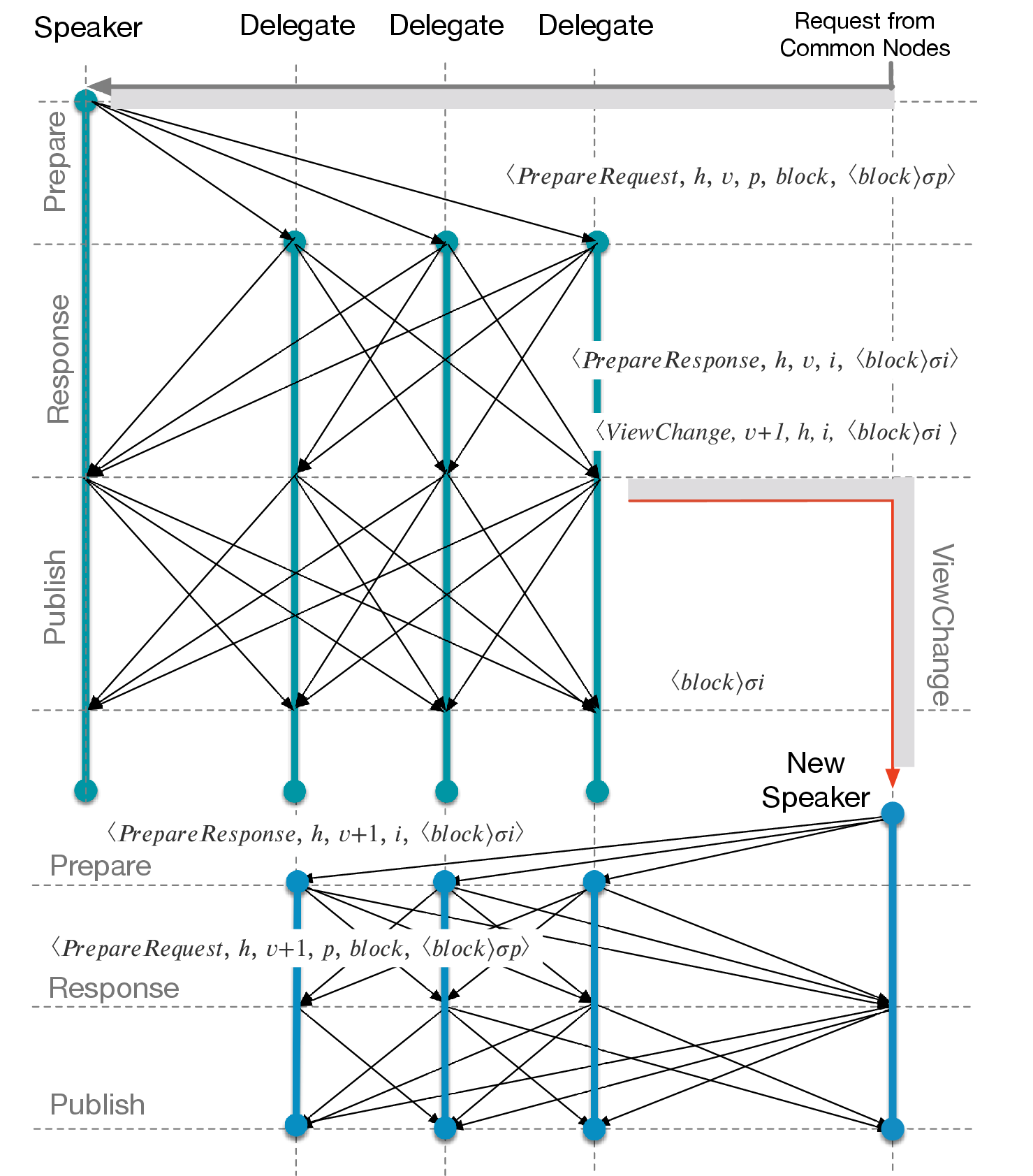}
\caption{dBFT Protocol}
\label{fig-NEODBFT}
\end{figure}

NEO system is a multi-participant flat topology system, containing the Byzantine nodes that can behave arbitrarily. Also, the network may fail to unconsciously or maliciously duplicate/deliver/delay the message, making the system out of order. Thus, the challenge is to guarantee that non-faulty nodes can execute the same operation with a consistent output. As a variant of PBFT, dBFT follows the same network assumption, where the nodes are not fully trusted and the network is partially synchronous. We provide an overview of the dBFT protocol in Fig.\ref{fig-NEODBFT} and the notations in Table.\ref{tab-notation} as the guideline.

\begin{table*}[!hbt]
 \caption{Notations of Participants}\label{tab-notation}
 \label{node}
  \centering
    \begin{tabular}[t]{cll}
    \toprule
    \textbf{ Symbol }  & \textbf{Item}  &\textbf{ Functionalities} \\  \midrule
     $\mathcal{N}$ &  Consensus group & participate the consensus procedure, each member is labeled by the index $i$, where $\{1,...,i\}\in \mathcal{N}$ \\  \midrule
     $p$ & speaker & the elected leader in the consensus group, where $p\in \mathcal{N}$  \\  \midrule
     $i$ & delegate & the replica in the consensus group, where $i\in \mathcal{N}$ \\ \midrule
     $f$ & Byzantine node &  an adversary who can arbitrarily behave during conse \\ \midrule
     $\mathcal{C}$ & client group &  synchronise the latest information, each client is instantiated as $c$, where $\{1,...,c,...\}\in \mathcal{N}$      \\ \midrule
     $\mathcal{V}/\mathcal{V}'$ & views &  the set of views and operated views, each view is instantiated as $v$   \\\midrule
     $\mathcal{O}/\mathcal{O}'$  & actions & the set of actions and operated actions, each action is instantiated as $o$   \\ 
    \bottomrule
    \end{tabular}
\end{table*}

\subsection{dBFT as SMR model}

In dBFT, states are replicated and broadcast across multiple nodes in a distributed environment. The dBFT consensus algorithm can be modelled as a state machine with replication. Each node that participates in the consensus acts as a replica to maintain the states for consistency. The delegates in the consensus group $\mathcal{N}$ are labelled by the indexes $\{0,...,|\mathcal{N}|-1\}$. To formally present the system, we abstract the NEO system based on State Machine Replication (SMR) model \cite{Schneider90}\cite{lynch1996}. The model presents the \textit{transition} of \textit{state}s\footnote{In this paper, we regard the term \textit{state} and \textit{view} as the same.} under specified \textit{action}s. An old state moves to a new state (also known as the final state) according to the given sequence of inputs and actions at a discrete-time tick. The \textit{action}s are marked with two notations containing $\mathsf{pre\_condition}$ and $\mathsf{effect}$, respectively, defining the achievability of inputs and the impact of the changes. Here we provide the definitions.

\begin{defi}[dBFT]
  The dBFT is a state machine denoted as a tuple $<\mathcal{V},\mathcal{V'},\mathcal{N},\mathcal{C}, \mathcal{O},\mathcal{O}',f ,v_0, net>$, and the dynamic transition state is defined as: \[
  f:\mathcal{V} \otimes \mathcal{N} \otimes \mathcal{C}\otimes  \mathcal{O}\xrightarrow{net} \mathcal{O}' \otimes  \mathcal{V'},  \]
  where $\mathcal{V}$ represents a set of states or views with the initial state $v_0$, $\mathcal{V'}$ is a group of new views after actions, $\mathcal{N}$ describes the consensus nodes while $\mathcal{C}$ represents non-consensus nodes or saying clients, $\mathcal{O}$ is a set of actions and $\mathcal{O}'$ indicates the operated actions under the transition function $f$, $net$ means the pair of messages targeting with its destination  $(m,i)$ under multicast channel network. 
\end{defi}

From the definition, dBFT can be regarded as a replicated state machine in an abstract presentation, in which the current view is transited into a new view under given actions among the participated consensus nodes. Each consensus node (delegates) in the system can be identified as $i\in \mathcal{N}$ while non-consensus node (client/common nodes) as $c\in \mathcal{C}$. The clients send requests to consensus nodes and wait for the replies after the consensus is completed. Since the speaker is selected in turns among delegates $\mathcal{N}$, each node can play the role of both \textit{speaker} role and \textit{delegate} as it does in traditional distributed systems. The delegate, as a replica/backup, receives the state/view from the speaker and maintains the state to keep the consistency. Another important component is the network in the system, and it could be regarded as Multicast Channel (MC). A message $m$ in the message space $\mathcal{M}$, where $m\in\mathcal{M}$, is transferred to its destination of a consensus node $i$, where $i\in\mathcal{N}$, which makes up a pair $(m,i)$. Each pair in the space forms the contents in MC, and we use \textit{net} to describe it. For simplicity, we use PREPARE and RESPONSE to represent the \textit{PrepareRequest} and \textit{PrepareResponse} in the source code.

The procedures of dBFT are presented in a specific description through the SMR model. Each procedure, containing several \textit{process}es executed by different \textit{role}s of members, represents a specified time interval. For \textit{role}, it contains three types of entities: \textit{client} $(c\in\mathcal{C})$, \textit{delegate} $(i\in\mathcal{N})$ and \textit{speaker} $(p\in\mathcal{N})$. Note that the speaker is elected from the delegates, so they belong to the same set. For \textit{process}, if a state gets updated under some actions with expected results, we regard it as a successful \textit{process}. Each process consists of two elements, the initial \textit{state} and the operational \textit{action}. The \textit{state} contains a serial descriptions on different events, such as $\mathsf{Blocksent}$, $\mathsf{ViewChanging}$, and $\mathsf{RequestSent}$. The \textit{action}, described in functionalities like $\mathsf{SEND}$, $\mathsf{REQUEST}$, \textit{etc}, is made up by a set of actions to finish the specified target.

To highlight the key procedures of dBFT, we capture two important elements described in SMR, namely \textit{view} and \textit{transition}. We regard \textit{view} as a collection of the messages, including proposals and responses in each process. We use $\textit{message}$ to describe the messages that are closely related to consensus. We emphasize several metrics such as the view $v$, or the block height $h$. The \textit{transition} contains a series of \textit{action}s in time sequence. Then, the messages are transferred under these actions to complete the transitions. Besides, some auxiliary functions are also provided in the following procedures. We introduce them in corresponding parts.


\subsection{Network Skeleton}

The nodes in the distributed system communicate with each other through a partially synchronous network. $m$, where $m\in\mathcal{M}$, represents the messages sent in the channel. $m$ includes the messages related to the transaction and block information. The message $m$ and its destination $i$, where $i\in \mathcal{N}$, are mapping to the pair $(m, i)$, denoted with \textit{net} as a component in the model. $<m>_{\sigma_i}$ means the message $m$ is signed by the delegate $i$. It is assumed that the \textit{net} records all messages exchanged in channels, providing a simulation environment for actions.

The MC network in the system provides  $\mathsf{SEND}$ and  $\mathsf{RECEIVE}$ actions for each direct channel between peers. $\mathsf{SEND}$ represents the actions executed at the origin, either by the client $c$ or the delegate $i$, and each of them is signed by the function $sig$ before disseminating. $\mathsf{RECEIVE}$ describes the actions to be executed by the destination node, to record the transferred messages when arriving at the destination. $\mathsf{MISBEHAVE}$ represents the messages that might be lost or duplicated in the channel, and SMR allows it to simulate the misbehaviour in \textit{net}.

For a normal transition, the $\mathsf{SEND}(m, \mathcal{N})_i$ action means that node $i$ broadcasts the messages to the destination set $\mathcal{N}$, \textit{i.e.},  all delegates. The $\mathsf{RECEIVE}$ action, randomly executed by one of the nodes in the destination set, reads the message and stores it into the memory pool. Note that, although the message can be sent to both $i$ and $c$,  we focus on the part of $i$ in MC. Equivalently, we only consider the events that relate to the consensus process.

\medskip   

   $\left.
    \begin{array}{ll}
    
     \textbf{State} \\
     net \subseteq \mathcal{M} \times 2^{\mathcal{N} },\,\,  \textrm{initial} \{\} \\
     net\_faulty \in Bool,\,\,   \textrm{initial}\,\, \textit{false} \\
     <m>_{\sigma_i}=sig(m,i),\,\,  \textrm{initial} \{\} \\

   \end{array}
   \right.$
  
   \medskip

   $\left .
    \begin{array}{ll}
    
     \textbf{Transitions} \\
     \texttt{SEND}(m, \mathcal{N})_i \\
     \quad \textrm{Eff}: net := net \cup (m, \mathcal{N})\\
     
   \end{array}
   \right.$
  
   \medskip

   $\left .
    \begin{array}{ll}     
    
     \texttt{RECEIVE}(m)_i\\
     \quad \textrm{Pre}: (m,\mathcal{N})\in net, i\in \mathcal{N} \\
     \quad \textrm{Eff}: net:=net - \{(m,\mathcal{N})\cup(m,\mathcal{N}-\{i\})\}\\
     
        \end{array}
   \right.$
  
   \medskip

   $\left .
    \begin{array}{ll}
     
     \texttt{MISBEHAVE}(m, \mathcal{N}, \mathcal{N}')\\
     \quad \textrm{Pre}:(m,\mathcal{N})\in net \land net\_faulty=true \\
     \quad \textrm{Eff}: net:=net - \{(m,\mathcal{N})\cup(m,\mathcal{N}')\}\\
     
   \end{array}
   \right.$
   \medskip

\subsection{Communication with Clients}

The communication with a client is modelled as a process, where the delegate $i$ receives the requests (transactions) from the client $c$ (lightweight/common nodes) and replies to the client after the corresponding execution. The request from the client $out_c$ is the input to the delegate $in_i$. The timestamp is remarked as $t$ while the operation is as $o$. The node failure in the equation represents the failure caused by nodes, including both malicious and unconscious ways.  The failure leads to the $retrans$ operation of the message. $retrans$ represents re-broadcasting the messages. Here are the parameters $o\in\mathcal{O}$, $c\in\mathcal{C}$, $i\in\mathcal{N}$, $r\in\mathcal{O'}$ and $m\in\mathcal{M}$.

For the normal transition without $node\_faulty$, the client $c$ sends the requests to the delegate $i$ by executing $\mathsf{REQUEST}$ and $\mathsf{SEND}$. The request adds the needed messages in the request to the $out_c$ and then invokes $\mathsf{SEND}$ to broadcast the request messages to all delegates. The delegate executes $\mathsf{RECEIVE}$ to add the request into its memory and executes the verification process to check the correctness of format, signature, hash value, \textit{etc}. After that, the client obtains the replied messages signed by the delegates via $\mathsf{REPLY}$, which includes the current view number $v$, timestamp $t$, delegate $i$, and client $c$. If the client is failed due to some reason, the $retrans_c$ will be set $true$ to repeatedly invoke $\mathsf{SEND}$.

    
     
     

\smallskip

   $\left.
    \begin{array}{ll}
    
     \textbf{State}\\
     v_c\in \mathcal{V},\,\, \textrm{initial}\; 0  \\   
     in_c \subseteq \mathcal{M},\,\, \textrm{initial}\; \{\} \\
     out_c \subseteq \mathcal{M},\,\, \textrm{initial}\; \{\} \\
     retrans_c \in Bool,\,\,  \textrm{initial false} \\
     node\_faulty_c \in Bool,\,\, \textrm{initial false} \\
     verify (m) \in Bool,\,\, \textrm{where}\,\, \{\forall i\in m, i =true\}\\
    
   \end{array}
   \right.$

\smallskip  
   
   $\left .
    \begin{array}{ll}

    \textbf{Transitions} \\
     \texttt{REQUEST}(o)_c \\
     \quad \textrm{Eff}: out_c=\{<\texttt{REQUEST},o,c,t, m_{\sigma_c}> \}\\
     \quad\quad\quad in_c := \{\} \\
     \quad\quad\quad retrans_c := false\\
     
        \end{array}
   \right.$
  
   $\left .
    \begin{array}{ll}
     
     \texttt{SEND}(m, \mathcal{N})_c\\
     \quad \textrm{Pre}: m\in out_c \land \neg retrans_c \\
     \quad \textrm{Eff}: out_c := out_c \cup \{<\texttt{REQUEST},o,c,t,m_{\sigma_c}>\ \} \\

   \end{array}
   \right.$

   $\left .
    \begin{array}{ll}
  
     \texttt{RECEIVE}(<\texttt{REQUEST},v,t,i,c,m_{\sigma_c}>)\\
     \quad \textrm{Eff}: verify ((<\texttt{REPLY},v,t,i,c,m_{\sigma_c}>) \\
     \quad \textrm{Eff}: in_i := in_i \cup \{<\texttt{REQUEST},o,c,t,m_{\sigma_c}> \} \\
     
        \end{array}
   \right.$
  
   $\left .
    \begin{array}{ll}
     
     \texttt{REPLY}(r)_c \\
     \quad \textrm{Pre}:out_c \neq 0 \land \exists \mathcal{N}: (|\mathcal{N}|>f \land \forall i \in \mathcal{N}:\\
     \quad\quad\quad \{ \exists v:(<\texttt{REPLY},v,t,i,c,m_{\sigma_i}> \in in_c)\}) \\
     \quad \textrm{Eff}: v_c := \{v|<\texttt{REPLY},v,t,i,c,m_{\sigma_i}> \in in_c \} \\
     \quad\quad\quad out_c=\{\} \\ 
     
        \end{array}
   \right.$

   $\left .
    \begin{array}{ll}
  
     \texttt{RECEIVE}(<\texttt{REPLY},v,t,i,c,m_{\sigma_i}>_c )\\
     \quad \textrm{Eff}: verify (<\texttt{REPLY},v,t,i,c,m_{\sigma_i} >_c) \\
     \quad\quad\quad  \textrm{if} \, true \\
     \quad\quad\quad  in_i := in_i \cup \{<\texttt{REPLY},v,t,i,c,m_{\sigma_i}>\} \\
     
        \end{array}
   \right.$

   $\left .
    \begin{array}{ll}

     \texttt{NODE-FAILURE}_c\\
     \quad \textrm{Eff}: node\_faulty_c := true\\
     
   \end{array}
   \right.$
  
   $\left .
    \begin{array}{ll}
         
     \texttt{SEND}(m, \mathcal{N})_c\\
     \quad \textrm{Pre}: m\in out_c \land  retrans_c \\
     \quad \textrm{Eff}: retrans_c := true\\
     \quad\quad\quad  out_c := out_c \cup \{<\texttt{REQUEST},o,c,t,m_{\sigma_c}> \} \\
     
   \end{array}
   \right.$

\subsection{Speaker and Delegates}

The delegates reply to the request from clients according to the consensus process. When the consensus starts, the nodes that participate in the consensus process decide their roles first by a deterministic formula $p=(h-v)\mod n$ to simulate a round-robin algorithm. The elected speaker executes and sends the \texttt{<Prepare>} message in the current round, and collects the \texttt{<Response>} messages from delegates. Both messages contain the information of view number $v$, block height $h$, node index $i$, and block payload. Note that, all the messages are signed by the sender.  When receiving messages, the receiver first verifies their validity. For the speaker and delegates, if the received message passes, the consensus proceeds as normal; if it fails or times out, the \textit{Viewchange} is launched to restart the consensus. A timer is embedded inside the node to avoid being blocked. The detailed mechanism of \textit{Viewchange} will be discussed later.

The parameters are set as follows. \textit{State} records and stores the initial parameters. \textit{Action} defines the behaviour of nodes. \textit{Auxiliary function} defines the functions used in consensus procedure, and each of them will be invoked for several times. \textit{Transition} refers to the state transition during the consensus. Three $\mathsf{SEND}$ actions separately represent the outputs towards the speaker $p$, delegates $i$ and clients $c$. $(v,t,h) \in \mathcal{V}$, $(i,j,p) \in \mathcal{N}$, $c\in\mathcal{C}$, $m \in \mathcal{M}$, $V \subseteq \mathcal{V'}$. The new view and current view will satisfy the equation as $\mathcal{V'}=\mathcal{V}\times (\mathcal{N} \rightarrow \mathcal{O}) \times (\mathcal{N} \rightarrow  \mathcal{V})$. The auxiliary function provides inter-generated functions used in the system. $primary(i)$ specifies the rules of speaker selection. $count$ describes the number of received signatures is more than certain threshold like $(2f+1)$. The notation $eql(v,i)$ means that the current view number of node $i$ is $v$, while $oper(m, \texttt{FUNCTION})$ represents the operations executed on $m$ by certain methods $\mathsf{FUNC}$ where $\mathsf{FUNC}=\{\texttt{REQUEST}, \texttt{REPLY}, \texttt{VIEWCHANGE},...\}$.

    
     
     


  
  \medskip

   $\left.
    \begin{array}{ll}
    
     \textbf{State} \\
     
     v_0 \in \mathcal{V}, \,\,\textrm{initial}\, v_0  \\
     out_i \subseteq \mathcal{M},\,\, \textrm{initial}\, \{\}  \\
     v_i \in \mathcal{V},\,\, \textrm{initial} \,0 \\
     node\_faulty_i \in Bool,\,\, \textrm{initial}\,\, false \\
     
   \end{array}
   \right.$
   
   \medskip
   
    $\left.
    \begin{array}{ll}
    
     \textbf{Action} \\
      \texttt{RECEIVE}(<\texttt{REQUEST},o,t,c,m_{\sigma_c}>_i) \\
      \texttt{RECEIVE}(<\texttt{PREPARE},v,h,p,m_{\sigma_p}>_i) \\
      \texttt{RECEIVE}(<\texttt{RESPONSE},v,h,j,m_{\sigma_j}>_i) \\
      \texttt{RECEIVE}(<\texttt{VIEWCHANGE},v,h,j,m_{\sigma_j}>)\\
      \texttt{NODE-FAILURE}_i\\
     
      \texttt{SEND-PREPARE}(m,v,h)_i \\
      \texttt{SEND-VIEWCHANGE}(v)_i\\

      \texttt{SEND}(m,\mathcal{N})_c \\
    \end{array}
    \right.$
   
\medskip

   $\left .
    \begin{array}{ll}
    
    \textbf{Auxiliary Function} \\
    primary(i) \equiv p =(h-v)\mod{\mathcal{N}}\\
    count() \geqslant (2f+1) \\
    eql(v,i) \equiv v_i=v \\
    oper(m,\texttt{FUNC}) \equiv m=<\texttt{FUNC}> \\
   \end{array}
   \right.$
   
\medskip

   $\left .
    \begin{array}{ll}
    
     \textbf{Transitions} \\

     \texttt{CONSENSUS}(<\texttt{REQUESR},m_{\sigma_c}>,<\texttt{REPLY},m_{\sigma_i}> ) \\

     \texttt{SEND}(m, \mathcal{N}-\{i\})_i \\
     \quad \textrm{Pre}: m\in out_i \land \neg oper(m,\texttt{REQUEST}) \\
     \quad\quad\quad \land \neg oper(m,\texttt{REPLY})\\
     \quad \textrm{Eff}: out_i := out_i-\{m\}\\

   \end{array}
   \right.$

   $\left .
    \begin{array}{ll}
     \texttt{SEND}(m, \{primary(i) \} )_i \\
     \quad \textrm{Pre}: m\in out_i \land oper(m,\texttt{REQUEST}) \\
     \quad \textrm{Eff}: out_i := out_i-\{m\}\\
     
     \texttt{SEND}(<\texttt{REPLY},v,t,c,i,r,m_{\sigma_i}>, \{c\} )_i \\
     \quad \textrm{Pre}: <\texttt{REPLY},v,t,c,i,r,_{\sigma_i}>\in out_i\\
     \quad \textrm{Eff}: out_i := out_i-\{ <\texttt{REPLY},v,t,i,c,r,m_{\sigma_i}>\}\\
     
   \end{array}
   \right.$

\subsection{View Change}

View change provides \textit{liveness} in protocols when the current view fails or the consensus is temporarily blocked. It guarantees the system continuously proceeds. In dBFT, the \textit{Viewchange} is triggered by \texttt{<Timeout>} from consensus nodes when states are inconsistent. View change effectively prevents the participated nodes from indefinitely waiting. When times out, the node $i$ will terminate accepting messages and restart the view with an increased index $v+1$. After receiving at least $2f+1$ \texttt{<ViewChange>} signatures, the nodes reset the parameters and launch a new view.

There is a timer in the system to calculate the time since a new view starts. If the agreement cannot be reached, probably due to the configured time slot, or the illegal transactions, the delegate nodes will broadcast \texttt{<VIEWCHANGE>}. Each of them has a counter. When receiving more than $\lceil\frac{2n}{3}\rceil$ signed messages, the consensus restarts to enter the next round and re-selects a speaker. Initially, the delegates enter the view with the number $v=0$, and it increases along with the proceeding rounds. If view change is successfully triggered, the protocol enters a new round with the view number $v+1$. The lifetime of a view is configured as $F(t)=2^{v+1}t^\star$ \cite{NeoWP}, where $t^\star$ is the default time interval of the block generation and $v$ is the current view number. Note that with iterated rounds, the waiting time will exponentially increase, thus frequent \texttt{<VIEWCHANGE>} is avoided which makes nodes obtain enough time to reach the agreement. The view $v$ remains valid until the agreement is achieved or the system is crashed. View change mitigates the influence of network delay.


\subsection{Consensus}

Without malicious nodes or frustrating delays, the algorithm will proceed with the normal operations. Note that no matter whether receiving the \texttt{<REQUEST>} message or not, the agreement can always be achieved. When the consensus launches, the participated consensus nodes first decide their roles via $p=(h-v)\mod n$. Each node has the opportunity to be selected as the speaker. The probability is equivalent for every node. The selected speaker executes the PREPARE and invokes the $\mathsf{SEND}$ to broadcast the \texttt{<PREPARE>} message to all other delegates. When the delegate $j$ receives the messages, s/he verifies their correctness. If passed, the delegate $j$ executes RESPONSE and invokes $\mathsf{SEND}$ to broadcast the \texttt{<RESPONSE>} message to peers (including the speaker). The consensus nodes collect the \texttt{<RESPONSE>} messages and verify them while waiting for sufficient (more than the threshold) messages. Once the agreement is reached on the current state, the system publishes the block. For simplicity, we employ the functions of \texttt{REQUEST}, \texttt{PREPARE} and \texttt{PUBLISH} to separately represent the \textit{PrepareRequest}, \textit{PrepareResponse} and \textit{tell (block)} in source code.

\medskip

   $\left.
    \begin{array}{ll}
    
     \textbf{Normal Case Consensus} \\
     
     \texttt{RECEIVE}(<\texttt{REQUEST},o,c,t,m_{\sigma_c}>_i )\\
     \quad \textrm{Eff}: \textrm{if}\, t=t_i \, \textrm{then}\\
     \quad\quad\quad out_i:= out_i \cup \{<\texttt{REPLY},v_i,t,i,c,m_{\sigma_i}> \} \\
     \quad\quad\quad  \textrm{else if}\, t>t_i,  \textrm{then} \\ 
     \quad\quad\quad in_i :=in_i \cup \{<\texttt{REQUEST},o,c,t,m_{\sigma_c}>\} \\
     \quad\quad\quad \textrm{if}\, i \neq p \, \textrm{then}\\
     \quad\quad\quad  out_i:= out_i \cup \{<\texttt{REQUEST},o,c,t,m_{\sigma_c}>\} \\

   \end{array}
   \right.$

   $\left.
    \begin{array}{ll}     
    
     \texttt{PREPARE}(h,v,m)_i \\
     \quad \textrm{Pre}: primary(i)=i \\
     \quad \textrm{Eff}: out_i := \{<\texttt{PREPARE},h,v,i,m_{\sigma_i}> \}\\
    
   \end{array}
   \right.$

   $\left.
    \begin{array}{ll}
    
     \texttt{SEND-PREPARE}(m,v,h)_i\\
     \quad \textrm{Pre}:i=p \,\land V(v,i)\,  \land\\ 
     \quad\quad\quad \exists \; o,c,t:(<\texttt{REQUEST},o,c,t,m_{\sigma_i}>\,\land\, m\in in_i) \, \land \\
     \quad\quad\quad \nexists \; (<\texttt{PREPARE},h,v,m_{\sigma_i}> \in in_i) \\
     \quad \textrm{Eff}: out_i:=out_i \cup \{<\texttt{PREPARE},h,v,i,m_{\sigma_i}>\}\\

   \end{array}
   \right.$

   $\left.
    \begin{array}{ll}
 
     \texttt{RECEIVE}(<\texttt{PREPARE},h,v,m_{\sigma_p}>_j) \,(i\neq j) \\
     \quad \textrm{Pre}: j\neq p \,  \land  \nexists \; o,i,t:(<\texttt{PREPARE},o,i,t,m_{\sigma_i}> \notin in_j)\\
     \quad \textrm{Eff}:  verify (<\texttt{PREPARE},h,v,i,m_{\sigma_i}> )\,\,  \textrm{if} \, \, true \\
     \quad\quad\quad  in_j:=in_j \cup \{<\texttt{PREPARE},h,v,m_{\sigma_i}>\}   \\

      \end{array}
   \right.$

   $\left.
    \begin{array}{ll}     
     
     \texttt{RESPONSE}(h,v,m)_j \\
     \quad \textrm{Pre}: primary(j)\neq i \\
     \quad \textrm{Eff}: out_j := \{<\texttt{RESPONSE},h,v,i,m_{\sigma_j}> \}\\
     
     \texttt{SEND-RESPONSE}\\
     \quad \textrm{Eff}:  out_j:=out_j \cup \{<\texttt{REPONSE},h,v,i,m_{\sigma_i}> \}\\
  
   \end{array}
   \right.$

   $\left.
    \begin{array}{ll}
   
     \texttt{RECEIVE}<\texttt{RESPONSE},h,v,m>\\
     \quad \textrm{Pre}: \nexists \; o,i,t:(<\texttt{RESPONSE},o,i,t,m_{\sigma_i}> \notin in_j)\\  
     \quad \textrm{Eff}: verify ( <\texttt{RESPONSE},h,v,m_{\sigma_i} > ) \\
     \quad\quad\quad  \textrm{if} \, true \\
     \quad\quad\quad in_j:=in_j \cup (<\texttt{RESPONSE},h,v,m_{\sigma_i}>)   \\
     
   \end{array}
   \right.$
   
    $\left.
    \begin{array}{ll}
    
     \texttt{PUBLISH(m,v,h)}_i\\
     \quad \textrm{Pre}: count(\texttt{RESPONSE})\geq 2f+1 \\
     \quad \textrm{Eff}:  v_i=v_0 \\
     \quad\quad\quad out_j:=out_j \cup \{<\texttt{PUBLISH},h,v,i,m_{\sigma_i} >\}\\
     
   \end{array}
   \right.$

\medskip

\subsection{Security Goals}

\noindent\underline{\textit{\textbf{Safe Behaviour.}}}
The safe behaviors, denoted as $\mathbb{S}$, are used to define the actions that ensure the BFT-style protocols normally operating. In contrast, the failure of the system means the agreement cannot be achieved. It might be caused by different ways, such as \textit{network failure} or \textit{node destruction}. It is well known that the threshold of the BFT system is $f \leq \floor{\frac{n}{3}}$, where $n$ is the committee size \cite{CastroL99}. dBFT follows this assumption.

\medskip


    

    $\left.
    \begin{array}{ll}  
    
    \textbf{State} \\
     v\in \mathcal{V}, \,\,\textrm{initial} \, v_0 \\
     in \subseteq \mathcal{O}\times \mathcal{N},\\
     out \subseteq \mathcal{O'}\times \mathcal{N} \\
     \exists i\in \mathcal{N}, node\_faulty_i \in Bool, \,\,\textrm{initial}\,\, false\\
     \exists i\in \mathcal{N}, net\_faulty \in Bool, \,\,\textrm{initial}\,\, false\\
     n\_faulty \equiv |\, \{i\,|\,node\_faulty_i=true  \}\,| \\
     \quad\quad \textrm{where}\, count(i)=f \\
   
    \end{array}
   \right.$
  
  
  \medskip
  
    $\left.
    \begin{array}{ll}
   
    \textbf{Transitions}(\textbf{if}\, f \leq  \floor{\frac{n}{3}} )\\
    \texttt{NODE-FAILURE} \\
    \quad \textrm{Eff}: node\_faulty_i = true \\
    \texttt{NETWORK-FAILURE} \\
    \quad \textrm{Eff}: net\_faulty = true \\
    \texttt{Execute}\\
    \quad \textrm{Pre}: <o,t,i> \in in \\
    \quad \textrm{Eff}: in := in - \{<o,t,i> \} \\
    \quad\quad\quad  out := out \cup \{r,t,i\} \\
    
   \end{array}
   \right.$
   
 \medskip

Based on the safe behavior $\mathbb{S}$, we then define the \textit{liveness} and \textit{safety}, and the consensus security of dBFT. The definition of liveness and safety \cite{alpern1987recognizing} can be formally established through a predicate $P(t)$, where $t$ represents time, and $t_{0}$ refers to the starting time. Set $x$ to be a process that is deadlock-free. 

\medskip
\noindent\underline{\textit{\textbf{Liveness.}}} Liveness stipulates that all processes eventually decide on a value, and the definition is stated as:

\begin{defi}[Liveness]\label{def-liveness}
  $\forall t_1 \geqslant t_0, \exists t\geqslant  t_1, t< \infty: P(t)=True$, where $P(t)$ represents $x$ stops waiting at the time $t$.
\end{defi}

More specifically, the system will never be blocked for an endless loop, which means the Timer $T=2^{v+k}t$ in the \textit{Viewchange} phase cannot be infinite, where $t<<\infty$. It is important to ensure that there always exists at least one agreement (process) to be achieved in a finite round before the timeout. Therefore, there exists one certain round which makes enough delegates receive sufficient (more than the threshold) signatures within the specified time. Consequently, we conclude the lemma of liveness in dBFT as follows:

\begin{lem} [Liveness]\label{lem-liveness}
The consensus achieves the \textit{liveness} property under $\mathbb{S}$ whenever it satisfies: For all states $v\in\mathcal{V}$, there always exists at least one state $v_i$ at a certain round in which the delegates could receive enough signatures (over the threshold $\lceil{\frac{2n}{3}\rceil}$) to reach the agreement. Formally, $\forall v \in\mathcal{V}$, $\exists v_i \in \mathcal{V}$, satisfying 
\[  \{\,T\,|\,T=2^{v+k}t^\star<<\infty \cap \textrm{count}(S_{v_i})>\lceil\frac{2n}{3}\rceil \}, \]
where $v_i$ is the state $i$ identified by a value, $t^\star$ is the timespan at a fixed value, $T$ is an accumulated timer along with the increased round $k$, following the equation $T=2^{v_i+k}t^\star$. $S_{v_i}$ represents the received signatures on the state $v_i$ from others, $count()$ is a counter to calculate the specific number, and  $count(S_{v_i})$ represents the number of received signatures on state $v_i$ from other consensus nodes.
\end{lem}

\smallskip
\underline{\textit{Safety}} stipulates that no processes decide on different values, and the definition is defined as:

\begin{defi}[Safety]\label{def-safety}
  $\forall t \geqslant t_0: P(t)=False$, where $P(t)$ represents $x$ is in STOP or a deadlock state at the time $t$.
\end{defi}

More specifically, this definition indicates the state can never be deadlocked on bad processes. Equivalently, the consensus cannot be blocked or forked by faulty nodes. In dBFT, it is required that the number of accumulated signatures from faulty nodes should be no more than $\floor{\frac{n}{3}}$ out of total nodes, so that the safety can be achieved in a correct state. Consequently, we conduct the lemma of safety as follows:

\begin{lem} [Safety]\label{lem-safety}
The consensus achieves the \textit{safety} property under $\mathbb{S}$ whenever it satisfies
\[count(S_{v
_i\gets f}) \leq \floor{\frac{count(S_{v_j\gets n})}{3}}, \] 
where $f$ means the faulty node and $n$ is the whole consensus node, $[state]\xleftarrow{}[nodes]$ defines the signatures' origination, $S_{v_i\gets f}$ are the received signatures on state $v_i$ from faulty nodes. $S_{v_j\gets n}$ are the received signatures from the whole consensus nodes. $count$ is the counter of accumulated signatures. $count(S_{v_i\gets f})$ means the number of received signatures on state $v_i$ from faulty nodes while $count(S_{v_j\gets n})$ on state $v_j$ from the whole consensus nodes.
\end{lem}

\noindent\underline{\textit{\textbf{Security}}}. Based on above definitions, we further define the security of the dBFT protocol as: 

\begin{defi}
   The dBFT protocol is \textit{secure} when it satisfies the \textit{liveness} and \textit{safety} property in the actual implementation. 
  \label{def-secure}
\end{defi}

More specifically, we can assume that there exists a set of cases, which contains all types of faults: $\textit{FAULTY}=\{s\in(1,2,...)\,|\,F_s\}$. We conduct the lemma of security as follows:

\begin{lem} \label{lem-secure}
The dBFT protocol is $secure$, when satisfying that  
\[ \textit{liveness} \,\land\textit{safety} = \textit{false}, \] 
under the case $s$,
where the case  $F_s$ is defined as  \[ \textit{FAULTY}=\{s\in(0,1,2,...)\,|\,F_s\}= \varnothing.\]
\end{lem}

\section{Security Analysis}
\label{sec-analysis}

Assume that there exists a case of faulty $F_0 \subseteq \textit{FAULTY} $ under $\mathbb{S}$. The state of $F_0$ contains the $faulty$ that is caused by either malicious nodes $f_i\in \mathcal{N}$ or network failure $f_{net}\subseteq {net}$. We can obtain that:

    \medskip
    $\left[
    \begin{array}{ll}
    
    node\_faulty_i\,||\,net\_faulty=true \\
    count(\textrm{faulty}) = f  \\
    count(\mathcal{N}) = n =3f+1 \\
    count(\textrm{honest}) = n-f  \\

   \end{array}
   \right.$
   \medskip
    
In initial, one of the faulty (or malicious) nodes $f_1$ is selected to be the speaker, and it spreads two conflicting blocks as messages $m_L$ and $m_R$, where the number of each type is equal to the other. After the \textit{Prepare} phase, half of the delegates receive $m_L$ while the others receive $m_R$.

     \medskip
     $\left.
     \begin{array}{ll}
    
     \texttt{SEND-PREPARE}(m_L,m_R,v,h)_i\\
     \quad \textrm{Pre}:f_1=p, m_L\in\mathcal{\mathcal{Q_L}},m_L\in\mathcal{\mathcal{Q_R}}, \mathcal{Q_L}\subseteq \mathcal{M}\\
     \quad\quad\quad \mathcal{Q_R}\subseteq \mathcal{M}, \mathcal{Q_L}\cup \mathcal{Q_R}=\mathcal{M},\mathcal{Q_L}\cap \mathcal{Q_R}=\mathcal{\varnothing}\\
     \quad\quad\quad node\_faulty_i\,||\,net\_faulty=true \\
     \quad\quad\quad count(m_L)=count(m_R) \\
     \quad \textrm{Eff}: out_i:=out_i \cup \{<\texttt{PREPARE},h,v,p,(m_L)_{\sigma_p}>\}\\
     \quad\quad\quad out_i:=out_i \cup \{<\texttt{PREPARE},h,v,p,(m_R)_{\sigma_p}>\}\\

     \texttt{RECEIVE}(<\texttt{PREPARE},h,v,(m_L)_{\sigma_p}>_i) \\
     \quad \textrm{Pre}: i=\{1,2,...,\frac{n-1}{2}\} \\
     \quad \textrm{Eff}:  in_i:=in_i \cup \{<\texttt{PREPARE},h,v,(m_R)_{\sigma_i}>\}   \\

     \texttt{RECEIVE}(<\texttt{PREPARE},h,v,(m_R)_{\sigma_p}>_j )\\
     \quad \textrm{Pre}: j=\{\frac{n+1}{2},...,n\} \\
     \quad \textrm{Eff}:  in_j:=in_j \cup \{<\texttt{PREPARE},h,v,(m_R)_{\sigma_p}>\}   \\
    
    \end{array}
    \right.$
    \medskip

Each delegate verifies the received messages and makes the \texttt{<RESPONSE>} message. Since these messages are half to half between two blocks, it results in the numbers of signatures on, respectively,  $m_L$ and $m_R$ are equal to each other.
    
     \medskip
     $\left.
     \begin{array}{ll}

     \texttt{SEND-RESPONSE}(h,v,m_L,i)_l,l\in\mathcal{N}\\
     \quad \textrm{Eff}:  out_i:=out_i \cup \{<\texttt{REPONSE},h,v,,i,(m_L)_{\sigma_i} >\}\\
   
     \texttt{SEND-RESPONSE}(h,v,m_L,i)_l,l\in\mathcal{N}\\
     \quad \textrm{Eff}: out_j:=out_j \cup \{<\texttt{REPONSE},h,v,j,(m_R)_{\sigma_j}> \}\\

     \texttt{RECEIVE}<\texttt{RESPONSE},h,v,m_L>\\
     \quad \textrm{Eff}: S_L=in_{l_L}:=in_l \cup <\texttt{RESPONSE},h,v,(m_L)_{\sigma_i}> \\
     
     \texttt{RECEIVE}<\texttt{RESPONSE},h,v,m_L>\\
     \quad \textrm{Eff}: S_R=in_{l_R}:=in_l \cup <\texttt{RESPONSE},h,v,(m_R)_{\sigma_j}>   \\
    
    \end{array}
    \right.$
    \medskip 
     
Since the signatures on different messages are equal, the consensus cannot be reached in the current round. When it times out, each node will disseminate \texttt{<VIEWCHANGE>} messages to restart the next round. A new round gets started when the node receives enough messages over threshold.

     \medskip
     $\left.
     \begin{array}{ll}
     
     \texttt{SEND-VIEWCHANGE}(h,v,m_L,l)_l',l'\in\mathcal{N}\\
     \quad \textrm{Pre}: count(S_L) = count(S_R),v=v+k(k=1)\\
     \quad\quad\quad m_N\in\mathcal{M} \subseteq{V'} \\
     \quad \textrm{Eff}:  out_l':=out_l' \cup \{<\texttt{VIEWCHANGE},h,v+k,m_N,i>_{\sigma_i} \}\\
    
     \texttt{RECEIVE}<\texttt{VIEWCHANGE},h,v+k,(m_N)_{\sigma_p}>)_l'\\
     \quad \textrm{Eff}:  \mathcal{V}=v_0  \\
     \quad\quad\quad S_N=in_{l'_N}:=in_l' \cup <\texttt{VIEWCHANGE},h,v,(m_N)_{\sigma_l'} >  \\
      
    \end{array}
    \right.$
    \medskip

At the new round, a new speaker is selected to disseminate \texttt{<PREPARE>} messages $m_N$. The honest nodes will accept $m_n$ while the faulty nodes abandon it.

    \medskip
     $\left.
     \begin{array}{ll}
     
    \texttt{SEND-PREPARE}(m_N,v,h)_i',i'\in \mathcal{N}\\
    \quad \textrm{Pre}: count(\texttt{VIEWCHANGE}) \geqslant 2f+1, primary(i')=p \\
    \quad \textrm{Eff}: out_i:=out_i \cup \{<\texttt{PREPARE},h,v,,p,(m_N)_{\sigma_p}>\}\\
     
    \texttt{RECEIVE}(<\texttt{PREPARE},h,v,m_N>_{\sigma_p})_i' \\
    \quad \textrm{Pre}: i'\in\mathcal{N} \cap i\in\mathcal{N} = \varnothing \\
    \quad \textrm{Eff}:  in_i':=in_i' \cup \{<\texttt{PREPARE},h,v,m_{\sigma_i}>\}   \\ 
    
    \end{array}
    \right.$
    \medskip

The faulty nodes refuse to accept $m_N$. Meanwhile, they fake the signatures on $m_R$ (denoted as $S_R'$) that they had not signed at the last round. The total number of newly faked $S_R'$ in the current round and already received $S_R$ in the last round is equal to the signatures $S_N$ on $m_N$ signed by honest nodes. Note that, the faulty nodes can be instantiated as different roles in different cases. Probably, one is the malicious node with \textit{fake}, while the other one is the network delay with \textit{delay}.

     \medskip
     $\left.
     \begin{array}{ll}

     \texttt{SEND-RESPONSE}(m_L,v,h)_i',i'\in \mathcal{N}\\
     \quad\quad\quad: node\_faulty =true \\
     \quad \textrm{Eff}: \underline{S_R'= fake_{f_i}(m_R)} \\
     \quad\quad\quad out_i':=out_i' \cup \{<\texttt{RESPONSE},h,v,p,(m_N)_{\sigma_p}>\}\\
     \quad\quad\quad out_i':= S_R \cup S_R' \\ 

     \texttt{SEND-RESPONSE}(m_L,v,h)_i',i'\in \mathcal{N}\\
     \quad\quad\quad: net\_faulty =true \\
     \quad \textrm{Eff}: S_R'= delay_{f_i}(m_R) \\
     \quad\quad\quad out_i':=out_i' \cup \{<\texttt{RESPONSE},h,v,p,(m_N)_{\sigma_p}>\}\\
     \quad\quad\quad out_i':= S_R \cup S_R' \\ 
     
    \end{array}
    \right.$
    \medskip

According to these failures, two exemplifications will be described in detail in the following section. Here, we can calculate the signatures on each different messages. The signatures on message $m_N$ is calculated as
\[ count(S_{N\gets {(n-f)}}) = n-f = 3f+1-f = 2f+1, \]
while the signatures on message $m_R$ (including valid $S_R$ and faked $S_R'$, totally denote as $S_f$) is as
   \begin{equation*}
   \begin{aligned}
    count(S_{R\gets f}) &= count(S_R+S_R') \\
          &= \frac{n+f+1}{2} = \frac{4f+2}{2} = 2f+1. \\ 
   \end{aligned}
   \end{equation*}
  Therefore, we can see that
   \[ count(S_{R\gets f}) = 2f+1 = count(S_{N \gets (n-f)}). \] 
   
\smallskip
From the above, the signatures of the new block $S_N$ signed by honest nodes is equal to $S_f$ by faulty nodes. Thus, the consensus is still being blocked. When times out by Timer=$2^{v+k}t^\star$, the nodes turn to the \textit{Viewchange} phase. Due to the configuration of protocol tolerance time, the parameter $t$ is close to failure within its settings. Based on such processes, we can induct the following theorems.

\begin{thm}
  The case $F_0$ cannot provide \textit{liveness} under the safe behaviors $\mathbb{S}$ in the dBFT consensus mechanism.
  \label{thm-liveness}
\end{thm}

\begin{prf} This can be proved by the contradiction. From the case $F_0$, we can see that
   \[ count(S_{L}) = count(S_{R}). \] 
   
The equation means the signatures on state $L$ from the malicious nodes $f$ are equal to that on state $R$ from the honest nodes $n-f$. Therefore, the consensus turns into the \textit{Viewchange}. At the next round, we can see that 
   \[ count(S_{R\gets f}) = count(S_{N\gets n-f}). \] 
   
The consensus is still being blocked at the new round because the malicious nodes hold the $2f+1$ signatures in total. It causes another conflict in the following round. The accumulated time $T$ by timer in the \textit{Viewchange} will roundly increase, which may leads to two results: 
   \begin{itemize}
       \item[-] the system crash due to exceeding the pre-configured tolerance time $T_0$, where $T>T_0$; or 
       \item[-] being blocked for an endless loop if there are no constraints on the tolerance time, where $\lim_{k \to\infty} T = \infty$.
   \end{itemize}
   
\smallskip
Intuitively, both cases \textit{\textbf{contradict}} to the conditions defined in Lemma \ref{lem-liveness}. Therefore, the dBFT protocol cannot satisfy the liveness, and it is consequently contradicts to liveness in Definition \ref{def-liveness}. Therefore, Theorem \ref{thm-liveness} is proved. 
\end{prf}

\smallskip
\begin{thm}
  The case $F_0$ cannot provide \textit{safety} under the safe behavior $\mathbb{S}$ in the dBFT protocol. 
  \label{thm-safety}
\end{thm}

\begin{prf} The theorem of safety can be proved by the contradiction. From the case $F_0$, we can see that

   \begin{align*}
    count(S_{R\gets f}) = count(S_{N\gets {n-f}}) 
   \nless & \floor{\frac{count(S_{N\gets n})}{3}}. \\
   \end{align*}
The equation means the number of received signatures on state $R$ from the malicious nodes $f$ are equal to that on state $N$ from the honest nodes $(n-f)$. Since no malicious nodes sign on state $N$, the number of signatures on $N$ from the honest nodes $(n-f)$ is equal to signatures from the total nodes $n\in \mathcal{N}$. Thus, it is no more than the threshold, which cannot satisfy the condition of safety in Lemma \ref{lem-safety}, and it is consequently contradicted to the definition of safety in definition \ref{def-safety}. Thus, Theorem \ref{thm-safety} is proved.
\end{prf}

\begin{thm}
  The dBFT consensus is not \textit{secure} under the safe behavior $\mathbb{S}$ in the NEO system. 
  \label{thm-secure}
\end{thm}

\begin{prf}
   This theorem can be proved in the contradiction. From Theorem \ref{thm-liveness} and Theorem \ref{thm-safety}, we can observe that $\{F_0\,|\,\textit{Liveness}=false\}$ and $\{F_0\,|\,\textit{Safety} =false \}$. As a result, it can be concluded that $\{F_0\,|\,\textit{Liveness} \land \textit{Safety} =false \}$. As $F_0 \subseteq \textit{FAULTY}$, from Lemma \ref{lem-secure}, we obtain that
    \[ \textit{FAULTY}=\{i\in(0,1,2,...)\,|\,F_i\} \neq \varnothing. \] 
    This indicates that the dBFT protocol is not \textit{secure} under the safe behaviour $\mathbb{S}$, and it is \textit{\textbf{contradicted}} to the security in Definition \ref{lem-secure}. Thus, Theorem \ref{thm-secure} is proved.
\end{prf}

\section{Vulnerability In Attacks}
\label{sec-example}
In this section, we provide two types of attacks as examples to demonstrate the correctness of our formal analysis for better understanding. We focus on the attack logic and corresponding security analysis. There are a total of seven nodes (align with the number of nodes set by the NEO foundation team), and the system is supposed to tolerate $f=(n-1)/3=2$ Byzantine nodes. However, we show that we can break the consensus safety with no more than two Byzantine nodes. 

\subsection{Breaking safety when $f=2$}

In this part, we focus on the attack model, implementation, and security analysis under the case of Byzantine behavior. The system is supposed to tolerate $f=(n-1)/3=2$ Byzantine nodes when $n=7$. We provide the details to break the consensus safety, and thus perform the double-spending attack, with two Byzantine nodes.

\smallskip
\subsubsection{Attack Model}
Byzantine behavior means a node can conduct arbitrary actions, including both malicious or honest actions. The vulnerability is due to malicious behaviors controlled by Byzantine nodes. We give the definition of this case. 

\begin{defi}[$\widehat{f=2}$]
    Assuming that there are totally 7 delegates in the committee, and two of them (including the speaker) are Byzantine nodes, \textit{i.e.}, $f=2$. The attack under such condition is denoted as $\widehat{f=2}$.
    \label{def-casef2}
\end{defi}

\subsubsection{Attack Logic}
The protocol of \textbf{Logic $f=2$} holds five phases in each round. At the end of each round, the algorithms trigger the \textit{Viewchange} to re-launch the view. There are totally two Byzantine nodes, one is elected as the speaker, denotes as $\mathcal{A}_1$, and the other one becomes one of the delegates, denotes as $\mathcal{A}_2$. We capture one slot to how the rounds process. Here is the implementation logic.

\smallskip
\textbf{Round $i$}
\begin{itemize}
    \item[-] \textit{step1:} The first step is to select the speaker among delegates. The principle is based on the rule of $p=(h-v) \mod n$ to rotate the authority. 
    \item[-] \textit{step2:} We assume that the speaker $\mathcal{A}_1$ is the Byzantine node, with the action: simultaneously send two proposals \texttt{<PrepareRequest>} on \textit{block1} to half of the delegates, and \textit{block2} to the other half of the delegates. 
    \item[-] \textit{step3:} After a short time, the delegates in the network receive the proposal from the speaker. Half of the delegates receive signatures on \textit{block1}, whereas the other half on \textit{block2} (the number is $\frac{n+1}{2}=\frac{n-1}{2}+1$). All of them make and send the \texttt{<PrepareResponse>} according to their received proposals. 
    Assuming that there exists another Byzantine node $\mathcal{A}_2$ among the delegates. S/he makes the response on \textit{block1}, but still collects signatures on \textit{block2} at the same time.
    \item[-] \textit{step4:} The delegate receives the responses from peers: half on \textit{block1} while the other half on \textit{block2} (the number is $\frac{n+1}{2}$).
    \item[-] \textit{step5:} We can see that two views have an equivalent number of signatures, and the system confronts the conflict between two different views (we call this situation as \textit{tie}). The system traps in \textit{tie} until times out. The \textit{Viewchange} is triggered and starts to proceed.
\end{itemize}

\smallskip
\textbf{Round $i+1$}
\begin{itemize}
    \item[-] \textit{step1:} After the \textit{Viewchange}, the speaker is re-elected through the rotation. If the new speaker turns to another Byzantine node, the round follows the same steps as the previous round. If the new speaker turns to an honest node, the steps continue as follows.
    \item[-] \textit{step2:} The honest speaker broadcasts a new proposal \texttt{<PrepareRequest>} on \textit{block3} to other delegates.
    \item[-] \textit{step3:} The delegates in the network receive the signatures on \textit{block3} and accordingly make the \texttt{<PrepareResponse>}. At that time, the Byzantine node $\mathcal{A}_2$ has already retained a half number of signatures on \textit{block2} in the previous round. Based on that, s/he fakes one more response on \textit{block2} with his own signature (now the number is $\frac{n+3}{2}=\frac{n+1}{2}+1$). 
    \item[-] \textit{step4:} The delegate receives the responses from peer delegates, but $\frac{n+3}{2}$ on \textit{block2} while $(n-f)$ on \textit{block3}. Here, we can see that $(n-f) = \frac{n+3}{2}$ when $n=2$ as an initial configuration.
     \item[-] \textit{step5:} The state still turns to \textit{tie} until times out. The \textit{Viewchange} is triggered and starts to proceed again. 
\end{itemize}

Note that \textbf{Logic $f=2$} gives two \textit{step sets} separately shown in Round $i$ and Round $i+1$. The algorithm following this paradigm falls into the conflict. 

\begin{figure}[t]
\centering
\includegraphics[width=0.46\textwidth]{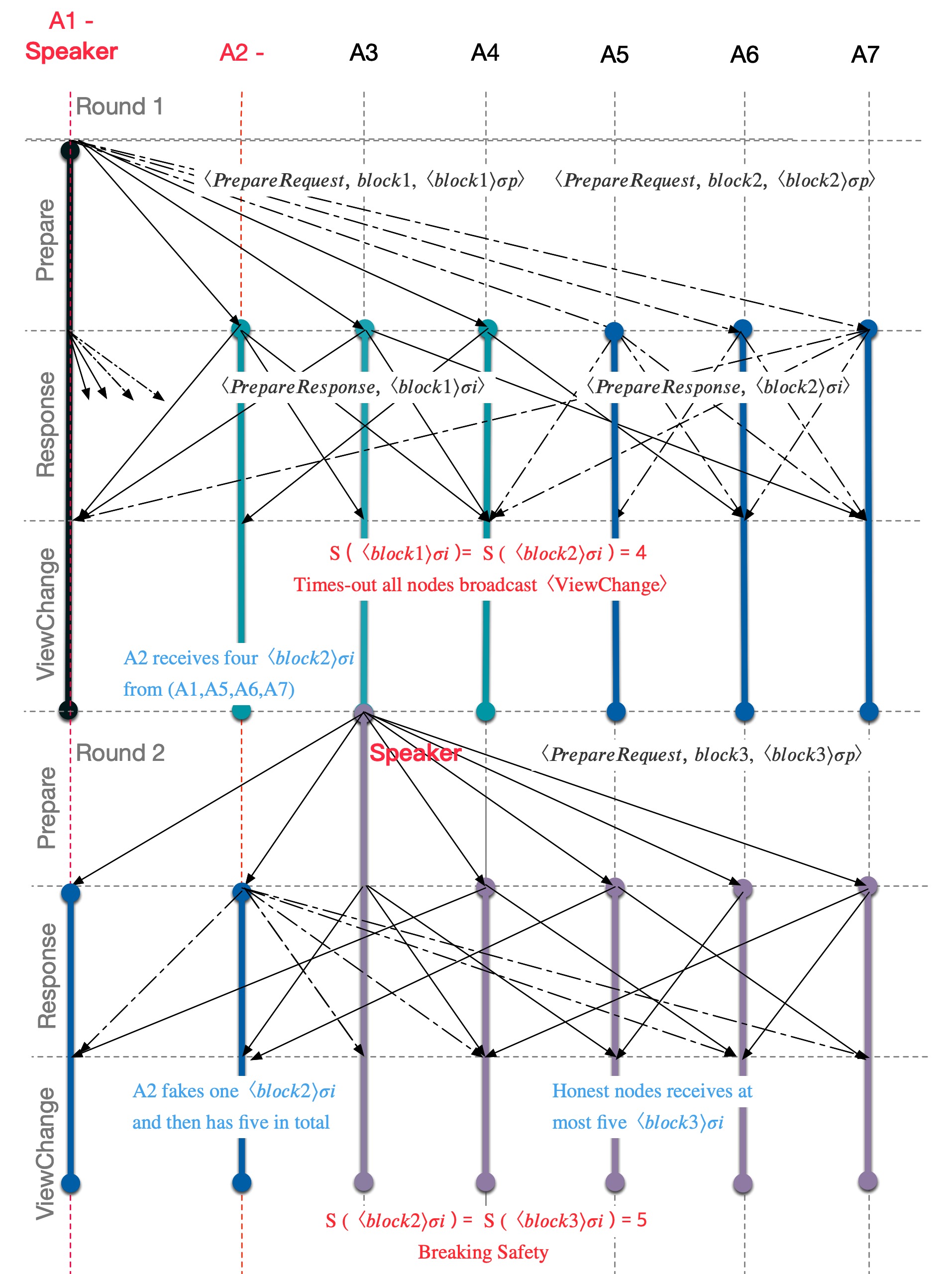}
\caption{Forks with $f=2$
}
\label{attack1}
\end{figure}

\begin{figure}[t]
\centering
\includegraphics[width=0.45\textwidth]{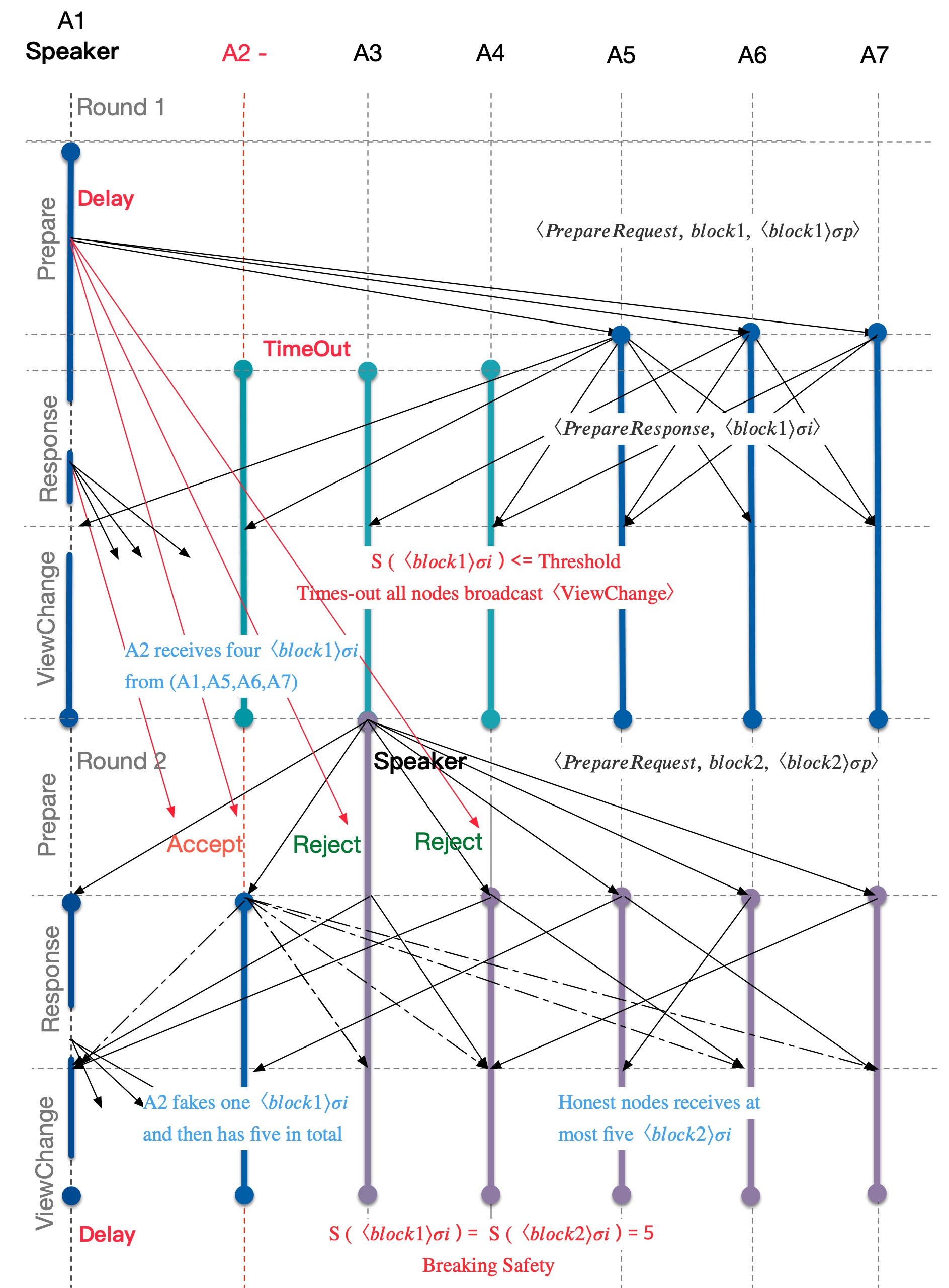}
\caption{Forks with $f=1$
}
\label{attack2}
\end{figure}

\smallskip
\subsubsection{Security Analysis} 
We provide a detailed security analysis to show why this case cannot satisfy the safety.

\begin{clm}[Breaking safety when $f=2$]
    The dBFT consensus cannot guarantee \textbf{safety} under the case of $f=2$.
    \label{clm-f2}
\end{clm}

\begin{prf}[$\widehat{f=2}$]:
Let $f=2$, and $A_1$ and $A_2$ are malicious, where $A_1$ is the speaker in the current round. The attack is presented in Fig.\ref{attack1}. In particular, $A_1$ generates two conflicting blocks, namely \textit{block1} and \textit{block2}. Then, $A_1$ sends $<PrepareRequest>_{block1}$ to $(A_2,A_3, A_4)$, and $<PrepareRequest>_{block2}$ to $(A_5,A_6,A_7)$.  $(A_2,A_3, A_4)$ will return the $<PrepareRespone>_{block1}$ message and $(A_5,A_6,A_7)$ return $<PrepareRespone>_{block2}$.

Now, there are four responses with signatures on both \textit{block1} and \textit{block2} in the network. The consensus cannot be reached, as at least $2f+1=5$ signed responses are required to make an agreement. In this case, the \textit{Viewchange} will be triggered, and every delegate resets their local states. In the new view, $A_1$ and $A_2$ behave normally, and the consensus on a block \textit{block3} prepared by another speaker can be reached. However, if the malicious node $A_2$ generates a newly signed response w.r.t. \textit{block2}, there will be five valid signed responses on this block, from $A_1$, $A_2$, $A_5$, $A_6$, and $A_7$. Thus, the condition to reach an agreement is achieved. In this way, there will be two agreements on the same round at the same height of the blockchain. It creates a fork in the system and breaks the safety.
\end{prf}

\subsection{Breaking safety when $f=1$}

In this part, we focus on the attack model, implementation, and security analysis in the case of package delay. Normally, the system is supposed to tolerate $f=(n-1)/3=2$ Byzantine nodes when $n=7$. But we show that we can break the consensus safety with only one Byzantine node.

\smallskip
\subsubsection{Attack Model}
Package delay means the proposals spread in the network are arbitrarily delayed for some reason. The vulnerability of dBFT, in this case, is caused by malicious nodes and package delays. We define it as follows. 
\begin{defi}[$\widehat{f=1}$]
    Assuming that there are a total of seven delegates in the committee, and one of them (the speaker) is a Byzantine node, \textit{i.e.}, $f=1$. The package is unpredictably delayed when sending the proposals. The attack under such a condition is denoted as $\widehat{f=1}$
    \label{def-casef1}
\end{defi}

\subsubsection{Attack Logic}
The protocol of \textbf{Logic $f=1$} holds five phases in each round. There is only one Byzantine node, denoted as $\mathcal{A}_2$. We assume that the package delay infects the transferred proposals, which leads to inconsistency as follows. 

\smallskip
\textbf{Round $i$}
\begin{itemize}
    \item[-] \textit{step1:} Select the speaker among delegates by the principle $p=(h-v) \mod n$. 
    \item[-] \textit{step2:} The honest speaker sends the proposal \texttt{<PrepareRequest>} on \textit{block1} to peers.
    \item[-] \textit{step3:} Due to the package delay, only half of the delegates (except $A_2$) receive the proposal in time, while the other half times out. The delegates who receive the proposal start to make and send the \texttt{<PrepareResponse>}.
    \item[-] \textit{step4:} Each delegate receives the responses from peers, and counts the number of received responses (the number is $\frac{n+1}{2}=\frac{n-1}{2}+1$, including the speaker).
     \item[-] \textit{step5:}  The state fails due to unsatisfying the threshold $\lceil\frac{2}{3}n\rceil$. The \textit{Viewchange} is triggered and starts to proceed.
\end{itemize}

\smallskip
\textbf{Round $i+1$}
\begin{itemize}
    \item[-] \textit{step1:} After the \textit{Viewchange}, the speaker is re-elected through rotation.
    \item[-] \textit{step2:} The speaker sends a proposal on \textit{block2} to peers. 
    \item[-] \textit{step3:} The delegates make the \texttt{<PrepareResponse>} according to \textit{block2}. At that time, the delayed proposal on \textit{block1} from the speaker is arrived. Normally the honest nodes will reject it, but the Byzantine node accept it and s/he fakes one more response on \textit{block1} with her/his own signature. The Byzantine node $\mathcal{A}_2$ has already received a half number of responses on \textit{block1} in the previous round (the number is $\frac{n+1}{2}$). Now the responses owned by $\mathcal{A}_2$ on \textit{block1} is $\frac{n+3}{2}=\frac{n+1}{2}+1$.
    \item[-] \textit{step4:} The delegate receives responses from peers, but $\frac{n+3}{2}$ of responses base on \textit{block1} while $(n-2)$ on \textit{block2}. 
     \item[-] \textit{step5:} As $\frac{n+3}{2}=(n-2)$ when $n=7$ is our initial configuration, the state still turns to \textit{tie} until times out. The \textit{Viewchange} is triggered and starts to proceed again. 
\end{itemize}

We noticed that although this case may not deterministically happen due to the unpredictable package delay. Once happen, it still leads to a serious monetary loss in the real life.

\smallskip
\subsubsection{Security Analysis} We provide the analysis as follows.

\begin{clm}[Breaking safety when $f=1$]
    The dBFT consensus cannot guarantee \textbf{safety} under the case of $f=1$.
    \label{clm-f2}
\end{clm}

\begin{prf}[case(f=1)]:
Let $A_2$ be the only malicious node, \textit{i.e.} $f=1$. The attack is presented in Fig.\ref{attack2}. In particular, assuming that $A_1$ is elected as the speaker in the current round, but the request on \textit{block1} sent by $A_1$ is delayed due to the package delay or network failure. As a result, $(A_5,A_6,A_7)$ have successfully received this request, and responded to $A_1$ with signed responses, whereas $(A_2, A_3, A_4)$ have not received \textit{block1} before the agreed time and no action has proceeded.

At this point, according to the dBFT protocol, the view will be changed when times out. The nodes start to run the \textit{Viewchange} algorithm, while $(A_2,A_3, A_4)$ still have not received neither the request on \textit{block1}, nor the signed responses from $(A_5,A_6,A_7)$.  After the view has been changed and a new block is proposed by a delegate, say $A_2$, an agreement on this new block is reached. Now, after the consensus on \textit{block2} is reached, the malicious $A_2$ receives the signed responses from $(A_1,A_5,A_6,A_7)$ on \textit{block1}. Then, $A_2$ can create her/his signed response on \textit{block1}. A valid agreement on \textit{block1} has been accordingly achieved, as five valid signed responses are collected. As a result, $A_2$ creates a fork in the blockchain, where two agreements are reached in the same round at the same height of the blockchain. Thus, it breaks the safety.
\end{prf}


\section{Recommended Fix}

As shown in the previous section, dBFT may confront a single-point of failure, \textit{i.e.}, trapping in conflicting states or inconsistent agreements even when $f=1$. This breaks both safety and liveness. The former is due to the fact that two inconsistent agreements are reached. It is possible to spend a coin multiple times in this case. The latter is a side effect of the former --- when inconsistent states exist in the system and there is no way to resolve them, such a situation will lead to a non-progression of the system.

Once the dBFT protocol is formally presented and fully understood by readers, the fix of the issue is straightforward. In particular, the most important observation is that, the absence of the \textit{Commit} phase after \textit{Response} in dBFT causes the problem. This is, in fact, a well-studied problem, such that it is possible to have a secure two-phase protocol for crash fault tolerance protocols, where the nodes can only be crashed. However, if several Byzantine nodes (with abilities to arbitrarily behave) are involved in a system, the \textit{Commit} phase becomes necessary \cite{CastroL99}\cite{FischerLP85}.

\begin{figure}[htbp]
\centering
\includegraphics[width=3.5 in]{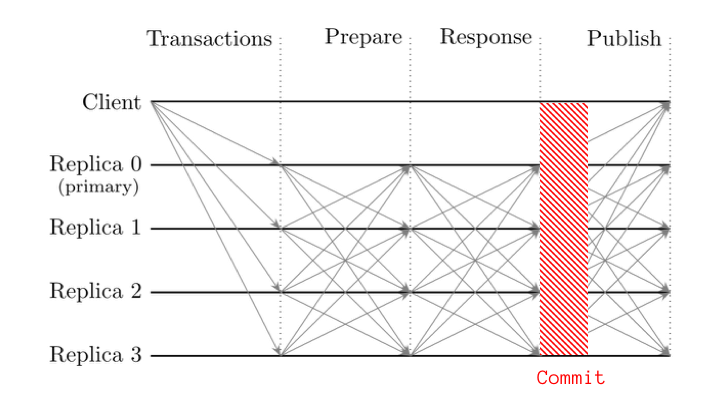}
\caption{dBFT with Fixed \textit{Commit} Phase}
\label{fig-no_commit_phase}
\end{figure}

The \textit{Commit} phase is used to check whether the received responded messages (\textit{w.r.t.} PREPARE message) from delegates exceeds the threshold (at least $2f+1$). If a node receives $2f+1$ signed responses from the delegates in the \textit{Response} phase, it commits the block by signing it together with the state information. If $2f+1$ COMMIT messages are received, the delegates will update the local state of the blockchain by including the block into it and broadcast the committed blocks to the network. In fact, this is a standard construction in classic BFT protocols, and it is proved to be secure in \cite{RahliVVV18}. However, due to the wrong adoption, dBFT fails in poor security. We have reported the issue to NEO \cite{neo320} and the corresponding fix (see Fig.\ref{fig-no_commit_phase}) is recommended on Github \cite{neo547}\cite{neo422}.

\section{Discussions}
\label{sec-discussion}

In this section, we conclude several related questions to extend the scope of our work from a thoughtful view and provide solid answers based on existing studies. 

\medskip
\textit{Q1.How does the issue come from?}
\smallskip

The dilemma confronted by dBFT, in fact, has been well studied for years. Back to the early research on such issues, the Two Generals Problem was proposed by Akkoyunlu et al. \cite{Akkoyunlu75} in the computer science field. The problem has been proved to be unsolvable in communication protocols \cite{PeaseSL80}. Moreover, the most prevailing Byzantine problem is also unsolvable in the case of network failure. VR \cite{OkiL88}, Paxos \cite{Lamport98}, and two-phase-commit (2PC) protocols are proposed to realize the multi-participants communications in distributed systems, but it can only solve the crash problems by non-human behaviors instead of the Byzantine problem by arbitrary human behaviors \cite{mohan1986transaction}. Other similar algorithms, including Zab, Zyzzyva, Kafka, and three-phase-commit (3PC) protocols \cite{vukolic2017rethinking}, cannot satisfy the requirements for practical applications, either. As analyzed before, dBFT is inherently a 2PC protocol, facing the same Byzantine problems regarding the consistency of consensus. 
  
\medskip
\textit{Q2.Will the issue be thoroughly solved?} 
\smallskip

The intuitive answer would be ``NO''.  Generally, the Byzantine problem emphasizes \cite{PeaseSL80}\cite{LamportSP82} unpredictable behaviours of distributed networks due to hardware errors, network congestion, and malicious attacks. The problem assists in blueprinting how non-fault nodes reach a consensus under unreliable networks. As discussed in \textit{Q1}, Byzantine problems confront theoretical limitations.

The FLP Impossibility theorem, proposed by Fisher et al. \cite{FischerLP85}, demonstrates the limitations of distributed processes. It proves that in an asynchronous system with multiple deterministic processes (where only one processor might crash), no solution exists for all processes to reach the agreement within a limited time. The FLP theorem proves the limits of finite time theoretical solution and as a result, researchers can only balance different factors in distributed systems to obtain a better trade-off, such as sacrificing certainty or increasing the processing time. Several blockchain systems evade FLP by setting weak network synchronization or variable confirmation periods, but still sacrifice one or more properties. 

The CAP theorem was proposed by Brewer et al. and proved by Gilbert et al. \cite{GilbertL02}. CAP represents the capital initial words of \textit{consistency}, \textit{availability} and \textit{partition tolerance}. Consistency means that all backups in a distributed system maintain the same data view at the same time. Availability refers to whether the cluster can handle or update the requests of clients when some nodes fail. Partition tolerance refers to the nodes in different partitions that can communicate with each other. The theorem implies that distributed systems cannot satisfy all three properties. Instead, it can only satisfy two of them at the same time. The CAP theorem makes researchers no longer pursue the perfect design with three characteristics at one time. A better trade-off among these properties becomes the target of the system design.

\medskip
\textit{Q3.How to mitigate the issue?} 
\smallskip

Since the BFT-style protocols have limitations with an upper bound (both security and performance), two technical routes are proposed when applied to blockchain systems. One is to follow the BFT design, but it requires taking careful consideration of configurations and edge conditions. Another is to leverage newly designed consensus mechanisms, such as PoX (Proof of something). This type of consensus transfers the selection process from the human to the substance such as work \cite{Satoshi08}, stake \cite{PPcoin12}, elapsed time \cite{chen2017security}, \textit{etc}. But it needs a strong network assumption to ensure that most blocks can be received. We conclude these two methods as follows.

\textit{Through BFT Consensus.} Various blockchain systems employ the BFT-style consensus, but instead, weaken the strict synchronous network assumption when adapting to Nakamoto’s consensus. The nodes need to know the latest blocks to avoid forks (the same as in the BFT assumption) but without setting an upper bound of the message delivery time (partial synchronous network assumption). The solutions include the dBFT of NEO, the BFT of Mir-BFT \cite{stathakopoulou2019mir}, the BFT of Hotstuff \cite{yin2019hotstuff}, the PoS+PBFT of Tendermint \cite{buchman2016tendermint}\cite{amoussou2018correctness}, the vote-based BFT of Algorand \cite{gilad2017algorand}, the Raft+PBFT of Tangaroa \cite{copeland2016tangaroa}, the variant PBFT of Thunderella \cite{pass2018thunderella}, PeerConsensus \cite{decker2016bitcoin}, ByzCoin \cite{kogias2016enhancing}, and Solida \cite{abraham2016solida}. 
All these solutions partially modify the original BFT protocols (mostly from PBFT \cite{CastroL99}) to fit their systems. dBFT, as one of the typical modifications, provides a warning example for the following project. The modification and combination of BFT-style protocols must primarily guarantee safety and liveness, and then consider the performance and other related properties. 

\textit{Through PoX Consensus.} 
Put forward by Dwork et al. \cite{DworkN92} and applied by Adam et al. \cite{adam02}, PoW (Proof of Work) was officially used in the Bitcoin system \cite{Satoshi08}. PoW requires the sender to work out the answer to a mathematical problem to prove that s/he has performed a certain degree of computational work. PoW ensures the decentralization and security of the system while perfectly integrating the currency issuance, circulation, and exchange. However, the increasing difficulty target of PoW causes useless power consumption and reduces the TPS of processing.  PoS (Proof of Stake) was proposed by Quantum Mechanic at the Bitcoin Forum and implemented in the project Peercoin \cite{PPcoin12}. The highest stakeholder will obtain the right of packaging, where stakes are evaluated by the time or the number of coins. It solves the problem of wasting computing power and decreases the confirmation time to reach an agreement. But the Matthew effect becomes significant with its negative influence. DPoS (Delegated Proof-of-Stake), applied in Bitshares \cite{bitshares18} and EoS \cite{Eos18}, is intrinsically a combination of the concept of PoS and BFT. The delegates of DPoS are selected according to their votes. Shareholders grant the rights to a certain number of delegates, and the selected ones are responsible for maintaining the chain. However, although PoX-based protocols possess the advantages of simplicity for decisions, they sacrifice the properties of efficiency and consistency \cite{bano2019sok}.

\medskip
\textit{Q4.What is the scope of the issue's impact?}
\smallskip

The most severe security problem in consensus mechanisms is the forks. We are going to clarify which types of the mechanisms might be vulnerable when confronting similar issues as dBFT. From one side, we observe that both BFT protocols and PoX protocols confront the threats by their centralized nodes, separately denoted as \textit{leader} in BFT and \textit{miner} in PoX. Inevitably, the authority of proposing blocks and the power of computing values are converged to a small group of supernodes in the real network. Therefore, the attacks targeting these nodes would be feasible. On the other side, we find differences between them. PoX protocols focus on how to avoid the centralization of a single miner, aiming to make the power distributed to multiple parties. Therefore, attacks like \cite{eyal2018majority} \cite{eyal2015miner} that utilize the advantages of miners are feasible. BFT protocols leverage deterministic mechanisms to elect the leader, and the selected leader can decide the final proposed blocks. The problem may happen \cite{abraham2018revisiting} during the procedure of informing other delegates for the message synchronization. Therefore, it is critical to guarantee the liveness and safety of the process. The issue of dBFT is caused by the removal of an essential phase. In contrast, the protocol with complete phases would not be influenced by such types of vulnerabilities. Therefore, 2PC protocols, either BFT-style or hybrid systems, need to carefully consider their designed protocols.

\medskip
\textit{Q5.What is the lesson learned from the issue?}
\smallskip

We summarize three major reasons to explain the issues of dBFT. Firstly, the issue is caused by the misusage of a mature protocol (PBFT) for a new project without thoughtful consideration. Simplifying a three-phase protocol to a two-phase protocol without any compensated solutions and auxiliary functions provides an educationally negative example for other projects when designing the protocol. The second reason is caused by the inappropriate changes in membership selection. Selecting the committee of delegates without any cost may lead to a high probability of corruption or bribe. Expanding the quorum size \cite{malkhi1998byzantine} and improving the cost of doing evil are beneficial to the trust of delegates. Thirdly, the network delay will impact multiple phases in a consensus protocol. Most of the systems lack a detailed description of the network assumption. We suggest taking related factors into consideration when configuring the protocol.

\section{Related Work}
\label{sec-relatedwk}
In this section, we provide several concepts surrounding this work, covering \textit{insecurities in permissioned blockchains} and \textit{formal treatment in blockchains}

\smallskip
\noindent\textbf{Insecurities in Permissioned Blockchains.} Existing attacks against blockchain consensus protocols are well-studied \cite{tsankov2018securify}\cite{zhang2019lay}\cite{peng2021security}\cite{ferrag2021performance}. We focus on the attacks and vulnerabilities in permissioned consensus mechanisms, which is the core idea of this paper. Permissioned blockchain is different from permissionless ones in its committee configurations. A permissionless blockchain has a dynamic unlimited committee in which a normal user can arbitrarily join and leave (e.g. PoW \cite{nakamoto2008bitcoin}). A permissioned blockchain requires a static and fixed-sized committee, just as it in BFT-style consensus \cite{malkhi1998byzantine}. This brings the risk of being attacked by such a committee.  Zhang et al. \cite{zhang2022frontruning} explored a frontrunning attack in PoA (proof of authority, \cite{de2018pbft}) protocols to break the legal turn of participated block proposers. Parinya et al.\cite{ekparinya2019attack} discovered the Clone attack to double-spend the victim transaction in PoA. Similarly, attacks in Hyperledger Fabric \cite{androulaki2018hyperledger} are severe due to the system's partially centralized committee \cite{brotsis2020security}. In this work, we explored the vulnerability in dBFT, which follows the same principle of permissioned blockchains. The insecurity in this work is caused by both incautious code errors in implementations and intrinsic principles of permissioned BFT settings.

\smallskip
\noindent\textbf{Formalization in Blockchain.}  The blockchain contains a suite of components that cover from lightweight clients to the underlying network. Defining a formal treatment for the entire blockchain system is difficult, so many researchers started from single components or simplified protocols. Garay et al. \cite{garay2015bitcoin} presented the earliest formal model of Bitcoin backbone protocol by summarizing its functional protocol and security properties (\textit{liveness} and \textit{persistence}). Later, they extended the research from underpinned protocol to embedded (PoW) difficulty adjustment mechanism \cite{garay2017bitcoin}, bounded-delay communication model \cite{pass2017analysis}, and adaptive dynamic participation model \cite{garay2020full}. Besides, Fitzi et al. \cite{fitzi2018parallel} studied the formal model for parallel chains. They explored the basic properties of throughput and security, as well as their balance off.  Myrto et al. \cite{arapinis2019formal} and Poulami et al. \cite{das2019formal} formally discussed the securities of digital wallets. They analyzed the state-of-the-art wallet solutions as well as related properties. Thomas et al. \cite{kerber2020kachina} dived into the security model of private smart contracts, thanks to the Universal Composition (UC) model. Li et al. \cite{li2022smart} formally analyzed the usage of smart contract towards security protocols. Mike et al. \cite{graf2020accountability} extended the study from permissionless blockchains (e.g., Bitcoin, Ethereum) to permissioned systems, especially for Hyperledger Fabric, by formally discussing their accountability. Our formal treatment differs from them by focusing on a long-lasting issue at the consensus level. We leverage the SMR model to simulate each procedure and accordingly induct the result of its insecurity.

\section{Conclusion}
\label{sec-conclusion}

NEO, as the pioneer of public blockchain projects around the world, confronts severe security threats. Our security analysis focuses on the core component of NEO, \textit{i.e.}, its dBFT consensus protocol. As a variant of PBFT, the dBFT protocol removes an essential \textit{Commit} process compared to the original protocol (PBFT), leading to deterministic forks under specified conditions. In this paper, we provide a comprehensive security analysis on such issues. Based on its source code, we employ the state machine replica model to formally present the protocol for better understanding. Then, we theoretically analyze the security of dBFT in terms of liveness and safety. The results prove the insecurity of the system under some special cases (when the Byzantine delegates are less than $\floor{\frac{n}{3}}$). We provide two instantiated attacks to show how the issues happen. In fact, the insecurity of removing the \textit{Commit} phase in BFT protocols is well known. This paper provides a study to revisit this issue, as a lesson learned from the already deployed and widely adapted consensus algorithm. Furthermore, we give a comprehensive discussion on the origin and impact of the issue to provide inspiration for future consensus designs.

\smallskip
\noindent\textbf{Acknowledgement.}
We appreciate the team members for their ever contributions: Jiangshan Yu (Monash University, Australia), Zhiniang Peng  (Qihoo 360 Core Security, China), Van Cuong Bui (Swinburne University of Technology, Australia), Yong Ding (Peng Cheng Laboratory, China).  A concise version (with the core idea of this work) has been published in Financial Cryptography and Data Security, 2020 (FC'20) \cite{fc20neo}.

\bibliographystyle{IEEEtran}
\bibliography{bib}

\begin{thebibliography}{10}
\providecommand{\url}[1]{#1}
\csname url@samestyle\endcsname
\providecommand{\newblock}{\relax}
\providecommand{\bibinfo}[2]{#2}
\providecommand{\BIBentrySTDinterwordspacing}{\spaceskip=0pt\relax}
\providecommand{\BIBentryALTinterwordstretchfactor}{4}
\providecommand{\BIBentryALTinterwordspacing}{\spaceskip=\fontdimen2\font plus
\BIBentryALTinterwordstretchfactor\fontdimen3\font minus
  \fontdimen4\font\relax}
\providecommand{\BIBforeignlanguage}[2]{{%
\expandafter\ifx\csname l@#1\endcsname\relax
\typeout{** WARNING: IEEEtran.bst: No hyphenation pattern has been}%
\typeout{** loaded for the language `#1'. Using the pattern for}%
\typeout{** the default language instead.}%
\else
\language=\csname l@#1\endcsname
\fi
#2}}
\providecommand{\BIBdecl}{\relax}
\BIBdecl

\bibitem{fc20neo}
Q.~Wang, J.~Yu, Z.~Peng, V.~C. Bui, S.~Chen, Y.~Ding, and Y.~Xiang, ``Security
  analysis on dbft protocol of neo,'' in \emph{International Conference on
  Financial Cryptography and Data Security (FC)}.\hskip 1em plus 0.5em minus
  0.4em\relax Springer, 2020, pp. 20--31.

\bibitem{NeoWP}
NEO, ``Neo whiteopaper,''
  \emph{\url{http://docs.neo.org/zh-cn/whitepaper.html}}, 2018.

\bibitem{Neo18}
E.~Zhang, ``Neo consensus,''
  \emph{\url{http://docs.neo.org/en-us/basic/consensus/consensus.html}}, 2018.

\bibitem{ferrag2018blockchain}
M.~A. Ferrag, M.~Derdour, M.~Mukherjee, A.~Derhab, L.~Maglaras, and H.~Janicke,
  ``Blockchain technologies for the internet of things: Research issues and
  challenges,'' \emph{IEEE Internet of Things Journal}, vol.~6, no.~2, pp.
  2188--2204, 2018.

\bibitem{dai2019blockchain}
H.-N. Dai, Z.~Zheng, and Y.~Zhang, ``Blockchain for internet of things: A
  survey,'' \emph{IEEE Internet of Things Journal}, vol.~6, no.~5, pp.
  8076--8094, 2019.

\bibitem{Eth14}
V.~Buterin, ``Ethereum website[eb/ol],'' \emph{\url{https://www.ethereum.org}},
  2018.

\bibitem{nakamoto2008bitcoin}
S.~Nakamoto, ``Bitcoin: A peer-to-peer electronic cash system,''
  \emph{Decentralized Business Review}, p. 21260, 2008.

\bibitem{vukolic2015quest}
M.~Vukoli{\'c}, ``The quest for scalable blockchain fabric: Proof-of-work vs.
  bft replication,'' in \emph{International workshop on open problems in
  network security}.\hskip 1em plus 0.5em minus 0.4em\relax Springer, 2015, pp.
  112--125.

\bibitem{vukolic2017rethinking}
------, ``Rethinking permissioned blockchains,'' in \emph{Proceedings of the
  ACM Workshop on Blockchain, Cryptocurrencies and Contracts}.\hskip 1em plus
  0.5em minus 0.4em\relax ACM, 2017, pp. 3--7.

\bibitem{CastroL99}
\BIBentryALTinterwordspacing
M.~Castro and B.~Liskov, ``Practical byzantine fault tolerance,'' in
  \emph{Proceedings of the Third {USENIX} Symposium on Operating Systems Design
  and Implementation (OSDI), New Orleans, Louisiana, USA, February 22-25,
  1999}, 1999, pp. 173--186. [Online]. Available:
  \url{http://doi.acm.org/10.1145/296806.296824}
\BIBentrySTDinterwordspacing

\bibitem{hyperledgerfabric}
``Hyperledger fabric,''
  \emph{\url{https://cn.hyperledger.org/projects/fabric}}, 2019.

\bibitem{hyperledgersawthooth}
``Hyperledger sawtooth,''
  \emph{\url{https://cn.hyperledger.org/projects/sawtooth}}, 2019.

\bibitem{androulaki2018hyperledger}
E.~Androulaki, A.~Barger, V.~Bortnikov, C.~Cachin, K.~Christidis, A.~De~Caro,
  D.~Enyeart, C.~Ferris, G.~Laventman, Y.~Manevich \emph{et~al.}, ``Hyperledger
  fabric: a distributed operating system for permissioned blockchains,'' in
  \emph{Proceedings of the Thirteenth EuroSys Conference}.\hskip 1em plus 0.5em
  minus 0.4em\relax ACM, 2018, p.~30.

\bibitem{NeoGit}
NEO, ``Neo github,'' \emph{\url{https://github.com/neo-project}}, 2018.

\bibitem{Schneider90}
\BIBentryALTinterwordspacing
F.~B. Schneider, ``Implementing fault-tolerant services using the state machine
  approach: {A} tutorial,'' \emph{{ACM} Comput. Surv.}, vol.~22, no.~4, pp.
  299--319, 1990. [Online]. Available:
  \url{https://doi.org/10.1145/98163.98167}
\BIBentrySTDinterwordspacing

\bibitem{lynch1996}
N.~A. Lynch, \emph{Distributed algorithms}.\hskip 1em plus 0.5em minus
  0.4em\relax Elsevier, 1996.

\bibitem{castro2001full}
M.~Castro, ``Practical byzantine fault tolerance (full),''
  \emph{\url{https://www.microsoft.com/en-us/research/wp-content/uploads/2017/01/thesis-mcastro.pdf}},
  pp. 1--172, 2001.

\bibitem{Sam2021sok}
W.~Sam \emph{et~al.}, ``\textrm{SoK}: Decentralized finance (defi),''
  \emph{arXiv preprint arXiv:2101.08778}, 2021.

\bibitem{vukolic2012quorum}
M.~Vukoli{\'c}, ``Quorum systems: With applications to storage and consensus,''
  \emph{Synthesis Lectures on Distributed Computing Theory}, vol.~3, no.~1, pp.
  1--146, 2012.

\bibitem{neo320}
``Discussion and improvement on dbft,''
  \emph{\url{https://github.com/neo-project/neo/pull/320}}, 2019.

\bibitem{neo547}
``Discussion and improvement on dbft,''
  \emph{\url{https://github.com/neo-project/neo/pull/547}}, 2019.

\bibitem{FischerLP85}
M.~J. Fischer, N.~A. Lynch, and M.~Paterson, ``Impossibility of distributed
  consensus with one faulty process,'' \emph{J. {ACM}}, vol.~32, no.~2, pp.
  374--382, 1985.

\bibitem{RahliVVV18}
V.~Rahli, I.~Vukotic, M.~V{\"{o}}lp, and P.~J.~E. Ver{\'{\i}}ssimo,
  ``Velisarios: Byzantine fault-tolerant protocols powered by coq,'' in
  \emph{{ESOP}}, 2018, pp. 619--650.

\bibitem{Ont18}
O.~Team, ``Ont consensus,''
  \emph{\url{https://github.com/ontio/ontology/tree/master/consensus/dbft}},
  2018.

\bibitem{neomaster}
``Neo source code on github,''
  \emph{\url{https://github.com/neo-project/neo/tree/master/neo}}, 2019.

\bibitem{alpern1987recognizing}
B.~Alpern and F.~B. Schneider, ``Recognizing safety and liveness,''
  \emph{Distributed computing}, vol.~2, no.~3, pp. 117--126, 1987.

\bibitem{neo422}
``Discussion and improvement on dbft,''
  \emph{\url{https://github.com/neo-project/neo/pull/422}}, 2019.

\bibitem{Akkoyunlu75}
E.~A. Akkoyunlu, K.~Ekanadham, and R.~V. Huber, ``Some constraints and
  tradeoffs in the design of network communications,'' \emph{SIGOPS Oper. Syst.
  Rev.}, vol.~9, no.~5, pp. 67--74, Nov. 1975.

\bibitem{PeaseSL80}
M.~C. Pease, R.~E. Shostak, and L.~Lamport, ``Reaching agreement in the
  presence of faults,'' \emph{J. {ACM}}, vol.~27, no.~2, pp. 228--234, 1980.

\bibitem{OkiL88}
\BIBentryALTinterwordspacing
B.~M. Oki and B.~Liskov, ``Viewstamped replication: {A} general primary copy,''
  pp. 8--17, 1988. [Online]. Available:
  \url{https://doi.org/10.1145/62546.62549}
\BIBentrySTDinterwordspacing

\bibitem{Lamport98}
L.~Lamport, ``The part-time parliament,'' \emph{{ACM} Trans. Comput. Syst.},
  vol.~16, no.~2, pp. 133--169, 1998.

\bibitem{mohan1986transaction}
C.~Mohan, B.~Lindsay, and R.~Obermarck, ``Transaction management in the r*
  distributed database management system,'' \emph{ACM Transactions on Database
  Systems (TODS)}, vol.~11, no.~4, pp. 378--396, 1986.

\bibitem{LamportSP82}
L.~Lamport, R.~E. Shostak, and M.~C. Pease, ``The byzantine generals problem,''
  \emph{{ACM} Trans. Program. Lang. Syst.}, vol.~4, no.~3, pp. 382--401, 1982.

\bibitem{GilbertL02}
\BIBentryALTinterwordspacing
S.~Gilbert and N.~A. Lynch, ``Brewer's conjecture and the feasibility of
  consistent, available, partition-tolerant web services,'' \emph{{SIGACT}
  News}, vol.~33, no.~2, pp. 51--59, 2002. [Online]. Available:
  \url{https://doi.org/10.1145/564585.564601}
\BIBentrySTDinterwordspacing

\bibitem{Satoshi08}
S.~Nakamoto, ``Bitcoin: a peer-to-peer electronic cash system,''
  \emph{\url{https://bitcoin.org/bitcoin}}, 2008.

\bibitem{PPcoin12}
S.~N. Sunny~King, ``\textrm{PPCoin}: Peer-to-peer crypto-currency with
  proof-of-stake,''
  \emph{\url{https://peercoin.net/assets/paper/peercoin-paper.pdf}}, 2012.

\bibitem{chen2017security}
L.~Chen, L.~Xu, N.~Shah, Z.~Gao, Y.~Lu, and W.~Shi, ``On security analysis of
  proof-of-elapsed-time (poet),'' in \emph{International Symposium on
  Stabilization, Safety, and Security of Distributed Systems}.\hskip 1em plus
  0.5em minus 0.4em\relax Springer, 2017, pp. 282--297.

\bibitem{stathakopoulou2019mir}
C.~Stathakopoulou, T.~David, and M.~Vukoli{\'c}, ``Mir-bft: High-throughput bft
  for blockchains,'' \emph{arXiv preprint arXiv:1906.05552}, 2019.

\bibitem{yin2019hotstuff}
M.~Yin, D.~Malkhi, M.~K. Reiter, G.~G. Gueta, and I.~Abraham, ``Hotstuff: Bft
  consensus with linearity and responsiveness,'' in \emph{Proceedings of the
  2019 ACM Symposium on Principles of Distributed Computing}.\hskip 1em plus
  0.5em minus 0.4em\relax ACM, 2019, pp. 347--356.

\bibitem{buchman2016tendermint}
E.~Buchman, ``Tendermint: Byzantine fault tolerance in the age of
  blockchains,'' Ph.D. dissertation, 2016.

\bibitem{amoussou2018correctness}
Y.~Amoussou-Guenou, A.~Del~Pozzo, M.~Potop-Butucaru, and S.~Tucci-Piergiovanni,
  ``Correctness and fairness of tendermint-core blockchains,'' \emph{arXiv
  preprint arXiv:1805.08429}, 2018.

\bibitem{gilad2017algorand}
Y.~Gilad, R.~Hemo, S.~Micali, G.~Vlachos, and N.~Zeldovich, ``Algorand: Scaling
  byzantine agreements for cryptocurrencies,'' in \emph{Proceedings of the 26th
  Symposium on Operating Systems Principles}.\hskip 1em plus 0.5em minus
  0.4em\relax ACM, 2017, pp. 51--68.

\bibitem{copeland2016tangaroa}
C.~Copeland and H.~Zhong, ``Tangaroa: a byzantine fault tolerant raft,''
  \emph{Tech. Rep}, 2016.

\bibitem{pass2018thunderella}
R.~Pass and E.~Shi, ``Thunderella: Blockchains with optimistic instant
  confirmation,'' in \emph{Annual International Conference on the Theory and
  Applications of Cryptographic Techniques}.\hskip 1em plus 0.5em minus
  0.4em\relax Springer, 2018, pp. 3--33.

\bibitem{decker2016bitcoin}
C.~Decker, J.~Seidel, and R.~Wattenhofer, ``Bitcoin meets strong consistency,''
  in \emph{Proceedings of the 17th International Conference on Distributed
  Computing and Networking}.\hskip 1em plus 0.5em minus 0.4em\relax ACM, 2016,
  p.~13.

\bibitem{kogias2016enhancing}
E.~K. Kogias, P.~Jovanovic, N.~Gailly, I.~Khoffi, L.~Gasser, and B.~Ford,
  ``Enhancing bitcoin security and performance with strong consistency via
  collective signing,'' in \emph{25th $\{$USENIX$\}$ Security Symposium
  ($\{$USENIX$\}$ Security 16)}, 2016, pp. 279--296.

\bibitem{abraham2016solida}
I.~Abraham, D.~Malkhi, K.~Nayak, L.~Ren, and A.~Spiegelman, ``Solida: A
  blockchain protocol based on reconfigurable byzantine consensus,''
  \emph{arXiv preprint arXiv:1612.02916}, 2016.

\bibitem{DworkN92}
\BIBentryALTinterwordspacing
C.~Dwork and M.~Naor, ``Pricing via processing or combatting junk mail,'' in
  \emph{Advances in Cryptology - {CRYPTO} '92, 12th Annual International
  Cryptology Conference, Santa Barbara, California, USA, August 16-20, 1992,
  Proceedings}, 1992, pp. 139--147. [Online]. Available:
  \url{https://doi.org/10.1007/3-540-48071-4\_10}
\BIBentrySTDinterwordspacing

\bibitem{adam02}
A.~Back, ``Hashcash - a denial of service counter-measure,''
  \emph{\url{http://www.hashcash.org/hashcash.pdf}}, 2002.

\bibitem{bitshares18}
``Bitshares: Delegated proof of stake consensus,''
  \emph{\url{https://bitshares.org/technology/delegated-proof-of-stake-consensus}},
  2018.

\bibitem{Eos18}
EoS, ``Eos: Bft-dpos consensus,''
  \emph{\url{https://github.com/EOSIO/Documentation/blob/master/TechnicalWhitePaper.md}},
  2018.

\bibitem{bano2019sok}
S.~Bano, A.~Sonnino, M.~Al-Bassam, S.~Azouvi, P.~McCorry, S.~Meiklejohn, and
  G.~Danezis, ``Sok: Consensus in the age of blockchains,'' in
  \emph{Proceedings of the 1st ACM Conference on Advances in Financial
  Technologies}, 2019, pp. 183--198.

\bibitem{eyal2018majority}
I.~Eyal and E.~G. Sirer, ``Majority is not enough: Bitcoin mining is
  vulnerable,'' \emph{Communications of the ACM}, vol.~61, 2018.

\bibitem{eyal2015miner}
I.~Eyal, ``The miner's dilemma,'' in \emph{2015 IEEE Symposium on Security and
  Privacy}.\hskip 1em plus 0.5em minus 0.4em\relax IEEE, 2015, pp. 89--103.

\bibitem{abraham2018revisiting}
I.~Abraham, G.~Gueta, D.~Malkhi, and J.-P. Martin, ``Revisiting fast practical
  byzantine fault tolerance: Thelma, velma, and zelma,'' \emph{arXiv preprint
  arXiv:1801.10022}, 2018.

\bibitem{malkhi1998byzantine}
D.~Malkhi and M.~Reiter, ``Byzantine quorum systems,'' \emph{Distributed
  computing}, vol.~11, no.~4, pp. 203--213, 1998.

\bibitem{tsankov2018securify}
P.~Tsankov, A.~Dan, D.~Drachsler-Cohen, A.~Gervais, F.~Buenzli, and M.~Vechev,
  ``Securify: Practical security analysis of smart contracts,'' in
  \emph{Proceedings of the 2018 ACM SIGSAC Conference on Computer and
  Communications Security (CCS)}, 2018, pp. 67--82.

\bibitem{zhang2019lay}
R.~Zhang and B.~Preneel, ``Lay down the common metrics: Evaluating
  proof-of-work consensus protocols' security,'' in \emph{2019 IEEE Symposium
  on Security and Privacy (SP)}.\hskip 1em plus 0.5em minus 0.4em\relax IEEE,
  2019, pp. 175--192.

\bibitem{peng2021security}
K.~Peng, M.~Li, H.~Huang, C.~Wang, S.~Wan, and K.-K.~R. Choo, ``Security
  challenges and opportunities for smart contracts in internet of things: A
  survey,'' \emph{IEEE Internet of Things Journal}, vol.~8, no.~15, pp.
  12\,004--12\,020, 2021.

\bibitem{ferrag2021performance}
M.~A. Ferrag and L.~Shu, ``The performance evaluation of blockchain-based
  security and privacy systems for the internet of things: A tutorial,''
  \emph{IEEE Internet of Things Journal}, 2021.

\bibitem{zhang2022frontruning}
x.~Zhang \emph{et~al.}, ``Frontrunning block attack in poa clique: A case
  study,'' in \emph{2021 IEEE International Conference on Blockchain and
  Cryptocurrency (ICBC)}.\hskip 1em plus 0.5em minus 0.4em\relax IEEE, 2022,
  pp. 1--3.

\bibitem{de2018pbft}
S.~De~Angelis, L.~Aniello, R.~Baldoni, F.~Lombardi, A.~Margheri, and
  V.~Sassone, ``Pbft vs proof-of-authority: Applying the cap theorem to
  permissioned blockchain,'' 2018.

\bibitem{ekparinya2019attack}
P.~Ekparinya, V.~Gramoli, and G.~Jourjon, ``The attack of the clones against
  proof-of-authority,'' \emph{arXiv preprint arXiv:1902.10244}, 2019.

\bibitem{brotsis2020security}
S.~Brotsis, N.~Kolokotronis, K.~Limniotis, G.~Bendiab, and S.~Shiaeles, ``On
  the security and privacy of hyperledger fabric: Challenges and open issues,''
  in \emph{2020 IEEE World Congress on Services (SERVICES)}.\hskip 1em plus
  0.5em minus 0.4em\relax IEEE, 2020, pp. 197--204.

\bibitem{garay2015bitcoin}
J.~Garay, A.~Kiayias, and N.~Leonardos, ``The bitcoin backbone protocol:
  Analysis and applications,'' in \emph{Annual International Conference on the
  Theory and Applications of Cryptographic Techniques}.\hskip 1em plus 0.5em
  minus 0.4em\relax Springer, 2015, pp. 281--310.

\bibitem{garay2017bitcoin}
J.~Garay \emph{et~al.}, ``The bitcoin backbone protocol with chains of variable
  difficulty,'' in \emph{Annual International Cryptology Conference
  (CRYPTO)}.\hskip 1em plus 0.5em minus 0.4em\relax Springer, 2017, pp.
  291--323.

\bibitem{pass2017analysis}
R.~Pass, L.~Seeman, and A.~Shelat, ``Analysis of the blockchain protocol in
  asynchronous networks,'' in \emph{Annual International Conference on the
  Theory and Applications of Cryptographic Techniques (EUROCRYPT)}.\hskip 1em
  plus 0.5em minus 0.4em\relax Springer, 2017, pp. 643--673.

\bibitem{garay2020full}
J.~A. Garay, A.~Kiayias, and N.~Leonardos, ``Full analysis of \textrm{Nakamoto}
  consensus in bounded-delay networks.'' \emph{IACR Cryptol. ePrint Arch.}, p.
  277, 2020.

\bibitem{fitzi2018parallel}
M.~Fitzi, P.~Ga, A.~Kiayias, and A.~Russell, ``Parallel chains: Improving
  throughput and latency of blockchain protocols via parallel composition,''
  \emph{Cryptology ePrint Archive}, 2018.

\bibitem{arapinis2019formal}
M.~Arapinis, A.~Gkaniatsou, D.~Karakostas, and A.~Kiayias, ``A formal treatment
  of hardware wallets,'' in \emph{International Conference on Financial
  Cryptography and Data Security (FC)}.\hskip 1em plus 0.5em minus 0.4em\relax
  Springer, 2019, pp. 426--445.

\bibitem{das2019formal}
P.~Das, S.~Faust, and J.~Loss, ``A formal treatment of deterministic wallets,''
  in \emph{Proceedings of the 2019 ACM SIGSAC Conference on Computer and
  Communications Security (CCS)}, 2019, pp. 651--668.

\bibitem{kerber2020kachina}
T.~Kerber, A.~Kiayias, and M.~Kohlweiss, ``Kachina-foundations of private smart
  contracts.'' \emph{IACR Cryptol. ePrint Arch.}, p. 543, 2020.

\bibitem{li2022smart}
R.~Li, Q.~Wang, Q.~Wang, and D.~Galindo, ``How do smart contracts benefit
  security protocols?'' \emph{arXiv preprint arXiv:2202.08699}, 2022.

\bibitem{graf2020accountability}
M.~Graf, R.~K{\"u}sters, and D.~Rausch, ``Accountability in a permissioned
  blockchain: Formal analysis of hyperledger fabric,'' in \emph{2020 IEEE
  European Symposium on Security and Privacy (EuroSP)}.\hskip 1em plus 0.5em
  minus 0.4em\relax IEEE, 2020, pp. 236--255.

\end{thebibliography}


\end{document}